\documentclass[preprint2]{aastex}

\slugcomment{submitted to ApJ}

\shorttitle{The star-formation history of the universe}
\shortauthors{M.Rowan-Robinson}

\begin{document}


\title{The star-formation history of the 
universe - an infrared perspective}


\author{Michael Rowan-Robinson}
\affil{Astrophysics Group, Blackett Laboratory, Imperial College of Science 
Technology and Medicine,
Prince Consort Road, London SW7 2BZ}



\begin{abstract}
A simple and versatile parameterized approach to the star formation history allows a 
quantitative investigation of the constraints from far infrared and submillimetre counts
and background intensity measurements.

The models include four spectral components: infrared cirrus (emission from interstellar
dust), an M82-like starburst,
an Arp220-like starburst and an AGN dust torus.  The 60 $\mu$m luminosity function
is determined for each chosen rate of evolution using the PSCz redshift data for
15000 galaxies.  The proportions of each spectral type as a function of 60 $\mu$m
luminosity are chosen for consistency with IRAS and SCUBA colour-luminosity relations, and
with the fraction of AGN as a function of luminosity found in 12 $\mu$m samples. The 
luminosity function for each component at any wavelength
can then be calculated from the assumed spectral energy distributions.  
With assumptions about the optical seds corresponding to
each component and, for the AGN component, an assumed dependence of the
dust covering factor on luminosity, the optical and near infrared counts can
be accurately modelled.  High and low mass stars are treated separately, since the
former will trace the rate of star formation, while the latter trace the
cumulative integral of the star formation rate.
 
A good fit to the observed counts at 0.44, 2.2, 15, 60, 90, 175 and 850 $\mu$m can be found 
with pure luminosity evolution in all 3 cosmological
models investigated: $\Omega_o$ = 1, $\Omega_o$ = 0.3 ($\Lambda$ = 0), and $\Omega_o$ = 0.3, 
$\Lambda$ = 0.7.

All 3 models also give an acceptable fit to the integrated background spectrum.
Selected predictions of the models, for example redshift distributions for
each component at selected wavelengths and fluxes, are shown.  The efect of
including an element of density evolution is also investigated.

The total mass-density of stars generated is consistent with that observed, in all
3 cosmological models.
\end{abstract}


\keywords{infrared: galaxies - galaxies: evolution - star:formation - galaxies: 
starburst - cosmology: observations}


\section{Introduction}

In this paper I investigate what source counts and the background radiation 
at infrared and submm wavelengths can tell us about 
the star formation history of the universe.  Madau et al (1996) showed how the ultraviolet
surveys of Lilly et al (1996) could be combined with information on uv dropout
galaxies in the Hubble Deep Field to give an estimate of the star formation history
from z = 0-4 (see also Madau et al 1998).
However these studies ignored what is already well-known from
far infared wavelengths, that dust plays a major role in redistributing the
energy radiated at visible and uv wavelengths to the far infrared.

Subsequent to the Madau et al analysis several groups of authors have argued that the 
role of dust is crucial in estimates of the star formation rates at high redshifts.
Rowan-Robinson et al (1997) derived a surprisingly high rate of star formation at z = 0.5-1 
from an ISO survey of the Hubble Deep Field (HDF). Subsequent ISO estimates by
Flores et al (1999) appear to confirm the need to correct for the effects of dust in estimates of 
star-formation rates.

Large extinction correction factors (5-10) 
were derived at z = 2-5 by Meurer et al (1997, 1999). Pettini et al 
(1998), Dickinson (1998) and Steidel et al (1999) found that estimates of star 
formation rates derived 
from HDF and other Lyman drop-out studies do indeed need to be corrected upwards by a 
substantial factor (2-7).  Thus the high star
formation rates deduced from the ISO-HDF data (Rowan-Robinson et al 1997) are not 
as out of line as at first appeared.

Many authors have attempted to model the star formation history of the universe ab initio
(eg Kauffmann et al 1993, Cole et al 1994, Pei and Fall 1995, Baugh et al
1998, Blain et al 1999c, Somerville et al 2000, Kauffmann and Haehnelt 2000).  There is an even longer history
of attempts to find empirical models for the evolution of the
starburst galaxy population (eg Franceschini et al 1991, 1994, 1997, Blain and Longair 1993,
Pearson and Rowan-Robinson 1996, Guiderdoni et al 1997, 1998, Dwek et al 1998, Blain et 
al 1999a, Dole et al 2000).
The study of Blain et al (1999a) attempts a more systematic approach
with a parameterized set of models but still treat the spectral energy distributions (seds)
of the galaxies in a simplistic manner.
In the present paper I report the results of a parameterized approach to the star formation
history of the universe, which allows a large category of possible histories to be
explored and quantified. The parametrized models can be compared with a wide range
of source-count and background data at far infrared and submillimtre wavelengths to
narrow down the parameter space that the star formation history can occupy.  The approach
is similar to that of Guiderdoni et al (1998) and Blain et al (1999a), 
but differs in key respects outlined below.

An important development for understanding the role of dust in the star formation history
of the universe has been the detection of the submm background at wavelengths from 
140-850 $\mu$m using COBE data (Puget et al 1996, Schlegel et al 1998, Hauser et al 1998, 
Fixsen et al 1998, Lagache et al 1999).  There is still disagreement about the actual value of the background
intensity by a factor of 2 at 850 $\mu$m and 2.5 at 140 $\mu$m. 

Franceschini et al (1997), Guiderdoni et al (1998), Blain et al (1999a), Dwek et al (1998) 
and Gispert et al (2000) have presented models to
account for this background radiation.  The models of Guiderdoni et al are able to
account for the 350 and 850 $\mu$m background, but fail to fit the observed background
at 140 and 240 $\mu$m, even
if the lower values of Fixsen et al are adopted.  The models of
Franceschini et al (1997) invoke a new population of heavily obscured high redshift
starbursts, designed to account for the formation of spiral bulges and ellipticals. 
In this paper
I test whether the submillimetre background and counts can be understood in terms of a single
population of evolving star-forming galaxies.  Blain et al (1999a) put forward a range of models 
which they argue are consistent with all available counts and background radiation.  I shall
argue below that, amongst other problems, their
most successful models involve an unphysical discontinuity in the star formation rate.
Of the two most successful models of Dwek et al (1998), one ('RR') is taken from an earlier
version of the present work, and the other ('ED') involves a spike in the star formation
history.

The layout of the paper is as follows.
In section 2, I introduce a
parameterized characterization of the star formation history which is
both simple and versatile.  Section 3 discusses the 60 $\mu$m luminosity function and
evolution rate as derived from the IRAS PSCz galaxy redshift survey.  Section 4 discusses the
assumed spectral energy distributions of galaxies as a function of luminosity, used to predict
the luminosity functions at other wavelengths.  Section 5 gives combined least squares fits to
the counts at 60 and 850 $\mu$m, and the background intensity at 140-750 $\mu$m in 3 
selected cosmological models.  Section 6 discusses the properties of the best-fitting
models at a wide range of wavelengths.  The effect of including 
some density evolution in an $\Omega_o$ = 1 model is discussed in section 7.  
Finally section 8 gives discussion and conclusions.

A Hubble constant of 100 $km/s/Mpc$ is used throughout.

\section{Parametrized approach to star formation history}

In this paper I present a parameterized approach to the problem, investigating a wide
range of possible star formation histories for consistency with counts and background
data from the ultraviolet to the submm.

The constraints we have on the star formation rate, $\dot{\phi_{*}}$(t) are that
\medskip

(i) it is zero for t = 0

(ii) it is finite at $t = t_{o}$

(iii) it increases with z out to at least z = 1 ( and from (i) must eventually decrease
at high z).
\medskip

A simple mathematical form consistent with these constraints is

\medskip

$\dot{\phi_{*}}(t)/\dot{\phi_{*}}(t_{o})$ = exp Q(1-$(t/t_{o}))$ $(t/t_{o})^P$  (1)

\medskip

where P and Q are parameters 

( P $>$ 0 to satisfy (i), Q $>$ 0 to satisfy (iii)).  I assume that $\dot{\phi_{*}}(t)$ = 0
for z $>$ 10.

Equation (1) provides a simple but versatile parameterization of the star formation history, 
capable of reproducing most physically realistic, single-population scenarios.  
Figure 1 shows models of type eqn (1)
(a) chosen to mimic the Pei and Fall (1995) models based on analysis of the history of heavy
element formation, and also the Cole et al (1994) semianalytic simulations of the star 
formation history in a cold dark matter (CDM) scenario, (b) chosen to be consistent with the strong 
evolution rate seen by Lilly et al (1996) in
the CFRS survey, (c) fitted to far infrared and submillimetre source-counts and
background intensity in different cosmological models (see below).  Models of
the form (1) will not, however, be able to reproduce the very sharply peaked scenario ('ED') of
Dwek et al (1998), or the two-population model of Franceschini et al (1997).  In the scenarios
modelled by eqn (1) the formation of bulges and ellipticals are part of an evolving spectrum of 
star-formation rates. 

The physical meaning of the parameters is as follows.  The exponential term describes 
an exponential decay in the star formation rate on a time-scale $t_{o}/Q$, which can
be interpreted as a product of the process of exhaustion of available gas for star formation
(a competition between formation into stars and return of gas to the interstellar
medium from stars) and of the declining rate of galaxy interactions and mergers at later
epochs.  This parameter is the same as that used in the galaxy sed models of Bruzual and Charlot (1993).
The power-law term represents the build-up of the rate of star
formation due to mergers of fragments which are incorporated into galaxies.  P measures
how steeply this process occurs with time.  The ratio P/Q determines the location of the peak in the
star formation rate, $t_{peak}$, since $t_{peak}/t_o = P/Q$.  

A very important assumption in the present work is that the star formation rate should vary
smoothly with epoch.  Several earlier assumptions have assumed, for mathematical
convenience, discontinuous changes of slope in the star formation rate (eg Pearson and
Rowan-Robinson 1996, the Blain et al 1999a 'anvil' models).  Such discontinuities are
highly unphysical and I have eliminated them in this work. 
The assumption above that $\dot{\phi_{*}}(t)$ = 0 for z $>$ 10 represents just such an unphysical
discontinuity in eqn (1) and I have investigated the effect of replacing the power-law
term in eqn (1) by

\medskip
$(t/t_{o})^{P} - (t_{i}/t_{o})^{P}$
\medskip

where  $t_{i}$  is the epoch at which star formation commences.  This would then constitute a third 
parameter in the models, but one which has very little effect on the results, provided $ P > 1, z(t_{i})>>1$.
The Blain et al (1999a) 'peak-4' and 'peak-5' models assume that there is no star-formation at z $<$ 4, 5,
respectively, and this constitutes a strong discontinuity within the observable range of redshifts, so
would again not be consistent with the philosophy adopted here.  The Blain et al (1999a) 'peak-10'
model is reasonable well approximated by (P,Q) = (1.9, 10.15).  

We might expect that the cosmological model could have a significant effect on the
relationship between predicted counts and predicted background intensity, since
the latter is sensitive to how the volume element and look-back time change with
redshift.

To test this I have explored models with (a) $\Lambda$ = 0, for which all the required 
formulae are analytic (specifically, here, models with $\Omega_o$ = 1, 0.3), and (b) k = 0, 
for which some of them are (specifically, $\Lambda$ = 0.7, $\Omega_o$ = 0.3).  Relevant formula 
are given in Appendix A.
  
\section{60 $\mu$m luminosity function and evolution derived from IRAS PSCz survey}

Given an assumed (P,Q) I then determine the 60 $\mu$m luminosity function, using the
IRAS PSCz sample (Saunders et al 1999).  I fit this with the form assumed by Saunders et al (1990)

\medskip
$\eta(L) = C_{*} (L/L_{*})^{1-\alpha} e^{-0.5[log_{10}(1 + L/L_{*})/\sigma]^{2}}$   (2)

\medskip
and find that the luminosity function parameters can be well approximated as follows:

\medskip
for $\Omega_o = 1$:

\medskip
$\sigma$ = 0.744-0.033 W

$log_{10}L_{*}$ = 8.50 + 0.05 W - 2 $log_{10} (H_{o}/100)$

\medskip
where  W = Q -  1.125 P

\medskip
for $\Omega_o = 0.3 (\Lambda = 0)$:

\medskip
$\sigma$ = 0.748-0.028 W

$log_{10}L_{*}$ = 8.46 + 0.05 W - 2 $log_{10} (H_{o}/100)$

\medskip
where  W = Q -  1.10 P

\medskip
for $\Lambda = 0.7 (k=0)$:

\medskip
$\sigma$ = 0.783-0.030 W

$log_{10}L_{*}$ = 8.39 + 0.05 W - 2 $log_{10} (H_{o}/100)$

\medskip
where  W = Q -  1.067 P.

\medskip
I have assumed fixed values for   $\alpha$ = 1.09, $C_{*}$ = 0.027 $(H_{o}/100)^{3}$.
It is not clear that any previous studies have correctly taken account of the need to change
the 60 $\mu$m luminosity function as the rate of evolution is varied.  The study of
Guiderdoni et al (1998) explicitly violates the known constraints on the 60 $\mu$m
luminosity function at the high luminosity end and as a result the models predict far too many
high redshift galaxies at a flux-limit of 0.2 Jy at 60 $\mu$m, where substantial redshift surveys have
already taken place.  Blain et al (1999a) state that they have determined the luminosity
function from consistency with the 60 $\mu$m counts, but this process does not automatically
give detailed consistency with existing redshift surveys.

We can also use the PSCz data to determine a range of consistency for (P,Q), using the
$V/V_m$ test.  The predicted uncertainty in $<V/V_m>$ for a population of n galaxies,
$(12n)^{-1/2}$ can be used to assign a goodness of fit for each set of (P,Q) values.
Fig 7 shows the locus of the best-fitting models in the P-Q plane (for which $<V/V_m>$ = 0.5),
together with the $\pm 1-\sigma$ region, for $\Omega_o$ = 1.

\section{Assumed infrared and submm spectral energy distributions}

To transform this 60 $\mu$m luminosity function to other wavelengths, we have to
make an assumption about the spectral energy distributions.  I have explored a variety
of assumptions about these seds: (a) a single sed at all luminosities representative 
of starbursts; (b) composite seds which are a mixture of four components, 
a starburst component, a 'cirrus'
component, an Arp 220-like starburst, and an AGN dust torus  (cf Rowan-Robinson and Crawford 1989, 
Rowan-Robinson and Eftstathiou 1993, Rowan-Robinson 1995), with the proportions of
each depending on 60 $\mu$m luminosity (Rowan-Robinson 1999a,b).  Neither of these approaches gave 
satisfactory results and it was not possible to find a simultaneous fit to all the far infrared
and submillimtre counts and background spectrum in any cosmological model.  Models with luminosity-
dependent composite seds could be
found which fit the far infrared and submillimetre counts and 850 $\mu$m background, but they 
failed to account for the 140-350 $\mu$m background.
Finally I have derived counts and background spectrum separately for each of the 
four components and then summed.  

Thus  

\medskip
$$N_{\nu}(S_{\nu}) = \sum_{i=1}^4 \int \eta_{60}(L_{60}) t_i(L_{60}) dL_{60} 
\int_0^{z(L_{60},S_{\nu},\nu)} (dV/dz) dz$$  (3)

where $lg L_{\nu} = lg L_{60} + lg [L_{\nu}/L_{60}]$ 

$= 6.18 + lg S_{\nu} + lg (\nu/\nu_{60}) + 2 lg D_{lum}(z) + k(z,\nu) +lg [\dot{\phi_{*}}(z)/\dot{\phi_{*}}(0)]$,

$L_{\nu}$, $L_{60}$ are in solar units, $S_{\nu}$ is in Jy, $D_{lum}$ is in Jy, and the k-correction  

$k(z,\nu) = lg [L_{\nu}(\nu(1+z))/L_{\nu}(\nu)]$.

\medskip
For small z, 
$$dN_{\nu}(S_{\nu})\over {\eta_{60}(L_{60})(L_{60}/S_{\nu})^{3/2}}$$ = $$\sum_i t_i(L_{60}) 
(L_{\nu}/L_{60})_i^{3/2}$$ (4).

This approach finally gave satisfactory fits to all
available data.  It also allows a correct determination of the redshift distribution
of each type separately as a function of wavelength and flux-density, and the
proportion of each type contribution to the counts at any flux-density or to the background.  

I have used the latest predictions
for infrared seds of these components by Efstathiou et al (2000), and added
near ir-optical-uv seds corresponding to an Sab galaxy (Yoshii and Takahara 1988) for the cirrus and an HII galaxy
(Tol 1924-416, Calzetti and Kinney 1992) for the starburst component, respectively.  The starburst model is a good fit 
to multiwavelength data for M82 and NGC6092 (Efstathiou et al 2000), and also to far infrared
and submillimetre data for luminous starbursts (Rigopoulou et al 1996).

The normalization
between far infrared and optical-uv components is determined by  L(60$\mu$m)/ L(0.8$\mu$m) = 0.15
for the cirrus component, 5.3 for the starburst component, and 49 for the Arp 200
component.  For the AGN component I assume that L(10$\mu$m)/L(0.1($\mu$m) = 0.3 (cf
Rowan-Robinson 1995) for the most luminous AGN, and that this ratio increases with decreasing 
luminosity to account for the fact that the mean covering factor is higher at lower
luminosities.  So lg {L(10$\mu$m)/L(0.1($\mu$m)} = -0.52 + 0.1*(14.0-log L60).

For the cirrus component I have, somewhat arbitrarily, divided the optical
sed into a contribution of young, high-mass stars (dominating at $\lambda \leq 0.3 \mu m$) and a
contribution of old, low-mass stars (dominating at $\lambda \geq 0.3 \mu m$) (see Fig 5).  
The former are assumed to trace the
star formation rate, and so participate in the strong evolution of the form eqn (1),
but the latter trace the cumulative star formation up to the epoch
observed, effectively a negative evolution with increasing redshift.  
This treatment, though approximate, allows a reasonable prediction of the K- 
and B-band counts.  The two components in Fig 5 can be modelled, assuming $L \propto M^3$,
blackbody seds with $T \propto L^{1/2}$, and with a Salpeter mass function,
with mass-range 0.1-1 $M_{\odot}$ for the low-mass star component, and 8-40 $M_{\odot}$
for the high-mass star component.      

The proportions of the four components (at 60 $\mu$m) as a function of luminosity (table 1) 
have been chosen
to give the correct mean relations in the S(25)/S12), S(60)/S(25), S(100)/S(60) and S(60)/S(850)
versus L(60) diagrams (Figs 2-4).  Where predictions are being compared with IRAS 12 $\mu$m
data, or (later) with ISO 15 $\mu$m or 6.7 $\mu$ counts, account is taken of the width of
the relevant observation bands by filtering with a top-hat filter of appropriate half-width.
(0.23, 0.26 and 0.16 at 6.7, 12, and 15 $\mu$m respectively).  Otherwise observations were assumed to be
monochromatic.
The relative proportion of the 60 $\mu$m emission 
due to AGN dust tori as a function of L(60) is derived from the luminosity functions
given by Rush et al (1993) (see Fig 10).
The resulting mean seds as a function of L(60) are shown
in Fig 6.  Luminosity functions at different wavelengths and the infrared bolometric luminosity
functions are shown in Figs 8-12.  There is good agreement with measured luminosity
functions at wavelengths from 0.44-850 $\mu$m.  The agreement with the luminosity function derived
at 60 $\mu$m from PSCz data is unsurprising, since the luminosity function parameters were determined
from these data.  The fit to the 850 $\mu$m luminsity function of Dunne et al (2000) is more
impressive, since the transformation from 60 to 850 $\mu$m is based only on choosing the
$t_i(L_{60})$ to give the correct ridge-line in Fig 4.  The fit to the non-Seyfert data
of Rush et al (1993) at 12 $\mu$m is also impressive, again arising only from the choice of
$t_i(L_{60})$ to be consistent with the colour-luminosity diagrams in Fig 3.  The fit to
the Seyfert luminosity function at 12 $\mu$m is not fortuitous, since the $t_i(L_{60})$ for the
AGN component were chosen to give this agreement.  It is impressive that luminosity functions
derived in the far infrared can fit the data at 0.44 $\mu$m (B band): the only freedom in the models
to fit the B-band luminosity functions and counts is the amplitude of the optical sed relative
to the far infrared.  

Clearly it will be
important to have submillimetre data for a wide range of normal and active galaxies to
test and improve these seds.  But the approach of using accurate radiative transfer models, 
with realistic assumptions about dust grains,
which have been verified with observations of known galaxies, seems superior to 
modelling the sed as a blackbody with power-law dust grain opacity in which
the dust temperature is treated as a free parameter (as in Blain et al 1999a).
The latter approximation can only be valid for rest-frame wavelengths greater than 60 $\mu$m, ie for
accurate prediction of counts and background intensities at wavelengths $>$ 200 $\mu$m.
Useful predictions can certainly not be made at 15 $\mu$m without explicit treatment of PAHs.
These criticisms do not apply to the studies of Guiderdoni et al (1998), Dwek et al (1998),
Xu et al (1998), Dole et al (2000), 
whose assumed seds are similar to those used here (but do not include the detailed
dependence on luminosity required for consistency with Figs 2-4).  

I have also assumed that the same luminosity evolution function should be applied to the whole
60 $\mu$m function, ie to AGN, 'cirrus' and 'starburst' components.  I have investigated
the effect of making the 
switchover in proportions of different types of component at a fixed luminosity,
so that there is in effect a strong increase in the proportion of galaxies that are starbursts  
(or contain AGN) with redshift.  However this did not permit a fit to all the available
data, so I have assumed that the Table 1 proportions relate to the luminosity at zero redshift
and that all these luminosities are subject to the evolution.  This is an approach which works, but 
models with more complex evolution than used here are clearly possible.

A substantial part of
the illumination of the cirrus component in spiral galaxies is due to recently formed
massive stars, part of whose light escapes directly from star-forming regions despite the
high average optical depth in these regions.  In the starburst models of Efstathiou et al (2000), 
this corresponds to the late stages of their starburst models.  
If the typical starburst luminosity was
greater in the past then the emission from interstellar dust in the galaxy would also be
correspondingly greater.  However the application of a uniform evolution rate to both the starburst
and cirrus components of the far infrared emission is not strictly self-consistent.  At low redshifts
the cirrus is illuminated by both low-mass and high-mass stars, with both contributing
comparable energy inputs.  At early epochs the illumination will be predominantly by
high-mass stars, since there would have been far fewer low-mass stars in the galaxy.  So the
evolution of the cirrus component will be complex, depending on the star-formation history,
the evolution of metallicity and the possibly varying dust geometry in the galaxy.  
The evolution of the seds of each component with redshift, particularly of the cirrus
component, may be an important factor in understanding the counts and background, but
is not treated here (see discussion in section 8).    

It is possible that the evolution of AGN differs from that
of starburst at z $>$ 3, but this will have little effect on the far infrared counts and
background (there could be a significant effect at 15 $\mu$m, which will be worth further study).

Elliptical galaxies are not treated explicitly, though their star formation rate must have been
much greater in the past than at present.  I am assuming that ellipticals are quiescent
starburst galaxies, that their star formation proceeded in much the same way as we
see in current live starbursts, and that their star formation episodes are part of the
evolution history quantified here.  We have to think of this history as a series of short-
lived fireworks taking place in different galaxies at different times.  Similarly this 
approach does not track the different spiral types separately, but only as a global average
at each epoch.

\section{Combined fits to 60, 175 and 850 $\mu$m counts, and 140-750 $\mu$m background, 
and determination of P,Q}

I can now predict the counts and background intensity at any wavelength and by
comparing with observed values, constrain loci in the P-Q plane.  To determine
(P,Q) for any given cosmological model, I combine the constraints found at 60 $\mu$m
from the PSCz (section 4 above) with constraints from (1) deep counts at 60 $\mu$m (50 mJy),
(2) the observed source-counts at 850 $\mu$m at 1 and 4 mJy, (3) the background
intensity at 140, 350 and 750 $\mu$m.  Figure 7 shows the constraints from the $V/V_m$ test and the 
850 $\mu$m counts for the $\Omega_o$ = 1 case.  The other constraints are less important in
specifying P and Q.  For all 3 cosmological models,
values of (P,Q) can be found which give a satisfactory fit to all the available data.
As emphasized above, this outcome does depend strongly on the assumptions made about the seds.

An important constraint on the models is that the total mass of stars produced in galaxies
should be consistent with the mass of stars observed, $\Omega_{*} = 0.003 \pm 0.0009 
h^{-1}$ 
(Lanzetta et al 1996), and that it should be less than the total density of baryons
in the universe, $\Omega_* \le \Omega_b = 0.0125 \pm 0.0025 h^{-2}$ (Walker et al 1991).  

We can calculate the total mass-density of stars from the 60 $\mu$m luminsity density using
eqn (7) of Rowan-Robinson et al (1997), modified to take account of the latest Bruzual 
and Charlot galaxy evolution models (Madau et al 1998, Rowan-Robinson 2000), from

\medskip
$\Omega_* = 10^{-11.13} \xi h^{-2} l_{60} (t_o/10 Gyr)$  (5)

\medskip
where $l_{60}$ is the luminosity density in solar luminosities per $Mpc^3$,

$\xi = \int_0^1 [\dot{\phi_*(t)}/\dot{\phi_*(t_o)} ] d(t/t_o)$,

and the assumed fraction of opt-uv light being radiated in the far infrared has been assumed to be 
$\epsilon = 2/3$.

Table 4 gives the values of (P,Q) which provide the best fit to the far infrared and submillimetre
counts and background, for each of the 3 cosmological models considered, and the corresponding
values of $l_{60}$, $\xi$ and $\Omega_*$, for an assumed age of the universe $t_o$ = 13 Gyr.
The predicted values of $\Omega_*$ are consistent with the oberved value in each case.
Estimating $\Omega_*$ from the young stellar component at 2800 $\AA$ or from the
K-band luminosity density (with an assumed mass-to-light ratio) also give consistent results
for an assumed Salpeter mass-function.

\section{Predicted counts and integrated background spectrum from uv to submm}
  
Figure 13-19 shows the predicted source-counts in the 3 selected cosmological models, at
850, 175, 90, 60, 15, 2.2 and 0.44 $\mu$m.  The agreement with observations at infrared
and submillimetre wavelengths is extremely impressive.  Although the fits at 850 and 60
$\mu$m have been ensured by the least-squares procedure for determining P and Q, the
fits at 175, 90 and 15 $\mu$ are simply a consequence of the assumed seds and the
choice of the $t_i(L_{60})$.  There is not much difference
between the predictions of the 3 cosmological models at 60-850 $\mu$m.  At 15 $\mu$m there
is a difference between the models in the predicted numbers of sources at fluxes below 100 $\mu$Jy.  
The proportion of AGN dust tori at 12 $\mu$ agrees well that the data of Rush et al (1993) 
(15$\%$ brighter than 0.4 Jy).  Fig 17 shows that
the proportion of AGN at 15 $\mu$m is reasonably constant (15-20$\%$) for fluxes brighter than 3 mJy, but 
is predicted to fall rapidly towards fainter fluxes.  

The fits to the counts at B and K are surprisingly good, though the model does not at present have ingredients capable 
of accounting for the very faint K- and B-band galaxy counts.  This could be provided
either by a measure of density evolution (see below) or by steepening the faint-end luminosity function 
at z $>$ 1, which could in either case be attributed to a population of galaxies that had
merged into present-day galaxies.  The B-band galaxy counts are dominated by the (negatively
evolving) low-mass star component for $lg_{10} S(0.44 \mu m) \geq$ -4.7 (Jy) ($B_{AB} \leq$ 20.7)
and by the  (positively evolving) high-mass star component at fainter fluxes.  In the I-band (0.8 $\mu$m) the
crossover is at 23.7 m.  The predicted median redshift for $I_{AB} \leq$ 22.5 is 0.49, in good agreement
with the results of the CFRS survey (Lilley et al 1995), but for B $\leq$ 22, the predicted median
z is 0.54, rather higher than the $\sim$0.3 observed by Colless et al (1993), suggesting that
we do not have quite the correct luminosity function or evolution for the blue galaxy population.

Fig 20-23 show redshift distributions at selected wavelengths and fluxes (complete output of the
models is available at http://icstar5.ph.ac.uk/ $\sim$mrr/countmodels).  The median redshift at
S(850 $\mu$m) $\geq$ 2 mJy (Fig 20) is predicted to be 2.25, significantly deeper than the prediction
for S(0.44 $\mu$m) $\geq$ 0.1 $\mu$Jy ($B_{AB} \leq$ 26.4 m.), 1.63 (Fig 23).  Only for $B_{AB} \leq$ 29
do we reach the same predicted median z as for 850 $\mu$m.  This shows the power of the 850 $\mu$m surveys
and also the difficulty there will be in identifying the sources detected.

Figure 24 shows the predicted integrated background spectrum for the 3 cosmological
models compared with the submillimetre observations summarized in Table 3, the optical and
near infrared estimates of Pozzetti et al (1998) and the 15 $\mu$m estimate of Sergeant et al
(2000).   Dwek et al (1998a) have argued
by extrapolation from longer wavelengths that the 100 $\mu$m background should lie in the range
5-34 $nW m^{-2} sr^{-1}$, but as this is not a true measurement of the background I have not used it here
(my models lie within the estimated range).
Dwek et al (1998b), Gorjian et al (2000)  and Wright and Reese (2000) have given estimates 
of the background at 2.2 and 3.5 $\mu$m, which appear to be a factor 2 or more higher than
estimates derived from summing deep galaxy counts (Pozzetti et al 1998).  This would seem to
imply either incomplete subtraction of foreground emissions or some additional background unconnected with
galaxies and I have not used these estimates.  

The models shown in Fig 24 are consistent with the observations, though the predictions of the $\Omega_o$ = 1
model are on the low side, while those of the $\Lambda$ = 0.7 model are on the high side.  
Figure 25 shows, for the $\Omega_o$ = 1
model, the contribution of the different sed types to the background. The dominant contribution is
from the cirrus component at most wavelengths, so the prediction is that more of the
energy from starbursts is deposited in the general interstellar medium of a galaxy than
is absorbed in the early stages close to the location of the massive stars.  This dominance
by the cirrus component at submillimetre wavelengths also implies that many of the detected
sources should turn out to be rather extended (kiloparsec scales rather than the more
compact scales expected for nuclear starbursts).  Table 5 gives the contributions of
the four components, and totals, to the background flux in the wavelength ranges 
0.1-5, 5-100 and 100-1250 $\mu$m.  The totals are somewhat lower than those given by
Gispert et al (2000), based on summing all claimed measurements of the background.

\section{Effect of including density evolution}

In the framework of hierarchical, bottom-up galaxy formation scenarios, like those
based on cold dark matter, we expect that galaxies form as the result of mergers
between smaller fragments.  Thus we might expect to see a higher total density of
galaxies as we look back towards the past.  To test whether attributing part of the
star formation history to density evolution has a major effect on the predicted
counts and backgrounds, I have considered, for the $\Omega_o = 1 $ case,
a simple modification in which the
comoving density of galaxies varies with redshift as

$\rho(z) = \rho(0) (1+z)^n$.

For n=1, this means that the comoving number-density of galaxies at redshifts 2, 5 and 10 (our
assumed cutoff redshift), is increased by a factor 3, 6 and 11 compared with the present.
This would represent a very substantial degree of merging of galaxies over the observable 
range of redshifts.

Since the background intensity depends on the product of the luminosity and density 
evolution rates, the background spectrum will be unaltered if we simply increase P by $2n/3$.
At low redshifts the counts will be unaffected if P is increased by $n/3$.
I find that for a combined density evolution model with n = 1 and luminosity evolution 
with P = 1.5, Q = 5.4 (and $\Omega_o$ = 1) the fits to the counts at 15-850 $\mu$m
hardly change over the observed range,
but the predicted background is raised by about 0.2 dex, giving
better agreement with observations for this cosmological model.  For the $\Lambda$=0.7
model the fit to the background would be significantly worsened for models consistent 
with the observed counts.

\section{Predicted confusion-limited performance of infrared and submillimetre telescopes}

The source-count models derived here can be used to compare the ultimate performance of different infrared
and submillimetre telescopes.  To do this in a simple and transparent way, I treat confusion in a
simplified but consistent manner.  I assume that the 5-$\sigma$ confusion limit falls at a
source-density of 1 source per 40 beams, defined to be circular apertures of radius

\medskip
$\theta_{conf} = 0.6 \lambda/D$

where D is the diameter of the telescope.
The corresponding source-density limit is denoted $N_{conf}$, per sq deg. 
The corresponding  confusion limit, and the fraction of sources in different redshift bins can then be estimated.  
The above assumption is valid for a population of sources with N(S) $\propto S^{-1.5}$ (Murdoch
et al 1973, Condon 1974).  The exact calculation depends slightly on the source-count slope.

Table 6 gives the wavelength, 2 $\theta_{conf}$, $N_{conf}$, $S_{conf}$, and redshift distribution
for my best $\Omega_o$ = 1 model, with (P,Q) = (1.2,5.4).

\section{Discussion and Conclusions}

(1) I have developed a parameterized approach to the star formation history, which is sufficiently
versatile to reproduce many proposed model histories.  The model assumes that the evolution of 
the star formation rate manifests itself as pure luminosity evolution.  I have stressed the
importance of ensuring that the assumed luminosity function is consistent with available
60 $\mu$m redshift survey data and that the assumed spectral energy distributions are realistic.
The observed far ir and submm counts and
background then provide strong constraints on the model parameters.  
The best fit for the $\Omega_o$ = 1 model is close to the star-formation history predicted
by the CDM scenario of Cole et al (1994), but is significantly less peaked than the
predictions of Pei and Fall (1995).  

The models consistent with infrared and submillimetre counts and backgrounds
tend to show a flat star formation rate from z = 1-3, consistent with the observed star
formation history derived from HDF galaxies, using photometric redshifts, and other uv, infrared and
submillimetre data (these will be discussed in a subsequent paper, Rowan-Robinson 2000).
The most striking difference from previous modelling work is the dominant role predicted for the
cirrus component at submillimetre wavelengths.  This is a consequence of the requirement that
all galaxies share a common star formation history.  The models presented here are by no means
unique.  If a very strongly evolving, heavily obscured population of starbursts is postulated, as
in the models of Franceschini et al (1997) and Guiderdoni et al (1998), then such a population
will affect only the submillimetre counts and background (and perhaps the faintest 60 $\mu$m
counts and redshift distributions).  Existing 850 $\mu$m surveys (Hughes et al 1998, Barger et al 1998,
Eales et al 1999) suggest that at least some of the submm galaxies may be luminous, opaque starbursts.
The prediction here, however, is that most will show a cooler sed, and greater spatial extent,
than expected for a typical M82- or A220-like starburst.

Areas requiring further work are (i) the need to consider the evolution of the
shape of the seds, particularly for the cirrus component, with redshift.  The increased
star-formation rate at earlier times would tend to make the dust temperature higher, but this
is partially offset by the much lower abundance of low-mass stars at earlier times. The
evolution of the metallicity may also affect the grain properties and hence the seds.
(ii) the AGN dust tori models require a further parameter, the orientation, and this may 
affect the AGN counts at optical wavelengths.
(iii) the tendency for the probability of finding an AGN component to increase with far
infrared luminosity is not fully reflected in the approach followed here.
(iv) it will be worthwhile to extend the predictions of these models to radio and X-ray
wavelengths.  




\appendix

\section{Cosmological formulae}

Collecting together the expressions for the cosmological observables, most of which are analytic functions, 
for the models consider here, we have, for (a) $\Lambda$ = 0 :

\medskip
the comoving radius

\medskip
r(z) =  2 $(1-\Omega_{o})^{1/2}[\Omega_{o} z + (\Omega_{o} - 2)((1+\Omega_{o}z)^{1/2}-1)]/\Omega_{o}^{2}(1+z)$,

\medskip
the luminosity distance

\medskip
$D_{lum}(z)$ =  $(1-\Omega_{o})^{1/2}$ (1+z) r(z) $(c/H_{o})$, 

\medskip
the comoving volume

\medskip
V(z) = 0.5 $(r(1+r^{2})^{1/2}$ + ln $(r + (1+r^{2})^{1/2})$  $(c/H_{o})^{3}$,

\medskip
and the age of the universe (in units of the Hubble time)

\medskip
$ H_{o} t(z) = [(1-\Omega_{o})(1+\Omega_{o} z)]^{1/2}/(\Omega_{o}(1 + z))$
- 0.5 ln $[(2-\Omega_{o} + \Omega_{o} z + 2 [(1-\Omega_{o})(1+\Omega_{o} z)]^{1/2} )/(\Omega_{o} (1+z))]$,

\medskip
$H_{o} t_{o} = (1-\Omega_{o})^{1/2}/\Omega_{o}$ - 0.5 ln$[(2-\Omega_{o}+2 (1-\Omega_{o})^{1/2})/\Omega_{o}]$.

\medskip
For (b), k = 0, all the relations are analytic except that for r(z), which is an elliptic integral
and has to be evaluated numerically.

\medskip
the comoving radius

\medskip
r(z) =  $$2 (1-\Omega_{o})^{1/2} \int_{\sqrt{(1/(1+z)}}^{1} [\Omega_o/(1-\Omega_o) + y^6]^{-1/2} dy$$,

\medskip
the luminosity distance

\medskip
$D_{lum}(z)$ =  (1+z) r(z) $(c/H_{o})$, 

\medskip
the comoving volume

\medskip
V(z) = $ r^3 (c/H_{o})^3 /3$,

\medskip
and the age of the universe (in units of the Hubble time)

\medskip
$ H_{o} t(z) = 2 sinh^{-1} [ ((1-\Omega_o)/\Omega_o)^{1/2} (1+z)^{-3/2}]/[3 (1-\Omega_o)^{1/2}]$,

\medskip
$H_{o} t_{o} = 2 sinh^{-1} [ ((1-\Omega_o)/\Omega_o)^{1/2} ]/[3 (1-\Omega_o)^{1/2}]$.





\clearpage

\figcaption[fig.eps]{
Examples of star formation histories of the form eqn (1):
(a) Fitted to those predicted by Pei 
and Fall (1995) (from the top, (P,Q) = (2.83, 8.15), (3.23, 10.25), 
(3.52, 11.92))
from an infall model (their Figs 1d, 3d, 4d), using dereddened quasar Lyman $\alpha$ absorber 
data, and
(dotted curve) the prediction by Cole et al (1994) derived from a CDM scenario, (P,Q) = (1.5, 5.).
(b) For $\Omega_o = 1$ models consistent with Lilly et al (1996) CFRS data, defined
to have $\dot{\phi_{*}}(1)/\dot{\phi_{*}}(0) = 2^{3.9}$.
From top at high z: (P,Q) = (1, 5.8), (2, 7.4), (3, 9.0), (5, 12.2).
(c) Best-fit models for far infrared and submm counts and background for
$\Omega_o = 1$, (solid curve, (P,Q) = (1.2,5.4)); $\Omega_o = 0.3$, (dotted
curve, (P,Q) = 2.1, 7.3)); $\Lambda = 0.7$, (broken curve, (P,Q) = (3.0, 9.0)).
The curves for (b) and (c) have been displaced upwards by +1 and +2 respectively.}

\figcaption[fig2.eps]{
S(100)/S(60) versus 60 $\mu$m luminosity for PSCz galaxies.  The solid line shows the 
trend with luminosity of the weighted average of the four component seds adopted.}

\figcaption[fig3.eps]{
S(25)/S(12) and S(60)/S25) versus 60 $\mu$m luminosity for PSCz galaxies. }

\figcaption[fig4.eps]{
S(60)/S(850) versus 60 $\mu$m luminosity for galaxies observed by Rigopoulou et al (1996),
Dunne et al (2000, open circles) and Fox et al (2000).  }

\figcaption[fig5.eps]{
Adopted spectral energy distributions for the four components adopted in this study: 
cirrus (with optical emission split into low-mass (broken curve) and high-mass
(dotted curve) stars), M82-starburst, A220-starburst (models from Efstathiou et al 2000), 
AGN dust torus (model from Rowan-Robinson 1995), showing assumed optical/ir ratio at
log L60 = 14 (upper curve) and 8.}

\figcaption[fig6.eps]{
Average sed as a function of 60 $\mu$m
luminosity, ranging from
$log_{10} (L_{60}/L_{\odot})$ = 8 to 14.  The relative proportions of the different
components, at 60 $\mu$m, are given in Table 1 and the average is calculated
using eqn (4) (AGN dust torus component excluded).}

\figcaption[fig7.eps]{
P-Q diagram, with loci for models fitting 60 $\mu$m counts and $V/V_m$ test (solid curves) and 
 850 $\mu$m counts (dotted curves).  Counts
loci correspond to 
log N850(4 mJy) = 2.95 $\pm$0.15 (Hughes et al 1998), and log N850(1 mJy) = 3.90 $\pm$0.15 (Blain et al 1999b).
The locus corresponding to lg N60(50 mJy) = 1.30 (Hacking and Houck 1987) is almost identical to
the 60 $\mu$m locus, with a slightly wider uncertainty ($\pm$ 0.28 in Q).   }

\clearpage

\figcaption[fig8.eps]{
Luminosity functions at 60 $\mu$m for the 4 spectral components.
Units of $\phi$ are $(Mpc)^{-3} dex^{-1}$, luminosity ($\nu L_{\nu}$) in solar units.
All luminosity functions are for $\Omega_o$ =1 model, $H_o$ = 100 $km s^{-1} Mpc^{-1}$.
Observed points are derived from PSCz data.}

\figcaption[fig9.eps]{
Luminosity functions at 850 $\mu$m for the 4 spectral components.  The data are from Dunne
et al (2000).   }

\figcaption[fig10.eps]{
Luminosity functions at 12 $\mu$m for the 4 spectral components.  Observed points
taken from Rush et al (1993) (filled circles: Seyferts, open triangles: non-Seyferts).}

\figcaption[fig11.eps]{
Luminosity functions at 0.44 $\mu$m for the 4 spectral components.  Data for quasars derived
from PG sample and for galaxies from Loveday et al (1992).}

\figcaption[fig12.eps]{
Bolometric luminosity functions for the 4 spectral components.  Note that for luminosities greater
than $10^{12.5} L_{\odot}$ there is an approximately equal contribution from M82-type
starburst, Arp220-like starbursts and AGN dust tori.}

\figcaption[fig13.eps]{
Integral source counts at 850 $\mu$m.   Data are from Hughes et al (1998), Eales et al (1998),
 Smail et al (1997), Barger et al (1999), Blain et al (1999c), Fox (2000).  
 The 3 models shown are, from bottom at faint fluxes, for $\Omega_o = 1$ and (P,Q) = (1.2, 5.4), 
 (solid curve), for $\Omega_o = 0.3$, (P,Q) = (2.1, 7.3) (dotted curve) and for $\Lambda$ = 0.7,
 (P,Q) = (3.0, 9.0) (broken curve).}

\figcaption[fig14.eps]{
Source counts at 175 $\mu$m. Data points are from Kawara et al (1998), Guiderdoni et al (1998),
Dole et al (2000).
Models as in Fig 13.}

\clearpage

\figcaption[fig15.eps]{
Source-counts at 90 $\mu$m.  Data from IRAS PSCz (triangles) and ELAIS (filled circles)
are from Efstathiou et al (2000).  Models as in Fig 13.}

\figcaption[fig16.eps]{
Source counts at 60 $\mu$m.   Data are from Lonsdale et al (1990) (at 0.2-10 Jy), Hacking and Houck 
(1987) (at 50-100 mJy), Rowan-Robinson et al (1991),
Gregorich et al (1995) (higher point at 50 mJy: see Bertin et al 1997 for discussion of possible
contribution of cirrus to Gregorich et al counts).  Models as in Fig 13.}

\figcaption[fig17.eps]{
Source counts at 15 $\mu$m.  Data from Oliver et al (1997, O),
Serjeant et al (2000, ELAIS), Elbaz et al et al (2000, E) 
Rush et al (1993, R, crosses), Verma (2000, V, filled circles).
Models as in Fig 13.  The lower dash-dotted line shows counts of the AGN dust torus
population for the $\Omega_o$ = 1 model.}

\figcaption[fig18.eps]{
Differential source counts at 2.2 $\mu$m.  Data from McCracken et al (2000).  
Models as in Fig 13.  }

\figcaption[fig19.eps]{
Differential source counts at 0.44 $\mu$m.  Galaxy data is from Metcalfe et al (1995), 
quasar data from Boyle et al (1988).  
Models as in Fig 13: Lowe dotted curve is prediction for quasars in $\Omega_o$ = 1 model.  
The long-dashed line shows the effects of including density evolution (see text) on the
$\Omega_o$ = 1 model.  }

\figcaption[fig20.eps]{
Predicted redshift distribution at 850 $\mu$m, $log_{10} S(Jy) = -2.7$. Bin centred at z = 5.25 
refers to z $>$ 5. }

\figcaption[fig21.eps]{
Predicted redshift distribution at 175 $\mu$m, $log_{10} S(Jy) = -1.0$.  }

\clearpage

\figcaption[fig22.eps]{
Predicted redshift distribution at 60 $\mu$m, $log_{10} S(Jy) = -0.7$.  Contribution from cirrus is shown dotted,
that from starbursts shown as broken histogram.  Observed points are derived from the FSSz
redshift survey of Oliver et al 1996.}

\figcaption[fig23.eps]{
Predicted redshift distribution at 0.44 $\mu$m, $log_{10} S(Jy) = -7.0$ (corresponding to $B_{AB}$ = 26.4).
Dotted and dash-dotted show contributions of high-mass and low-mass stars, broken curve
shows contribution of starburst component.  }

\figcaption[fig24.eps]{
Predicted spectrum of integrated background for same models as Fig 9.
Data from Fixsen et al (1998)
(far ir and submm), Pozzetti et al (1998) (opt and uv), Serjeant et al (2000) (15 $\mu$m).}

\figcaption[fig25.eps]{
Predicted spectrum of integrated background for $\Omega_o = 1$ model,
showing contribution of the different components.   The contribution of
the Arp220-like starbursts is less than 0.01 $nW m^{-2}$ at all wavelengths.
The upper solid curve at long wavelengths shows the effect of including density evolution
(see text).}







\clearpage

\begin{table}
\caption{$t_i(L_{60})$, proportion of different sed types at 60 $\mu$m, as function of $L_{60}$.}
\begin{tabular}{crrrrr}
& & & & \\
$lg_{10} L_{60}/L_{\odot}$ & cirrus & M82-starburst & A220-starburst & AGN dust torus\\
& & & & \\
8.0 & 0.99 & 0.001 & 0.5e-10 & 0.0091\\
9.0 & 0.80 & 0.1988 & 0.5e-7 & 0.0012\\
10.0 & 0.50 & 0.4999 & 0.5e-4 & 0.5e-4\\
11.0 & 0.10 & 0.85 & 0.05 & 0.19e-4\\
12.0 & 0.1e-3 & 0.75 & 0.25 & 0.25e-6\\
13.0 & 0.1e-7 & 0.75 & 0.25 & 0.25e-8\\
14.0 & 0.1e-12 & 0.75 & 0.25 & 0.25e-10\\
\end{tabular}
\end{table}


\begin{table}
\caption{Observed source-counts}
\begin{tabular}{crrrr}
& & & \\
wavelength ($\mu$m) & $log_{10}$ S(Jy) & $log_{10} N(>S)$ & reference\\
& & & \\
15 & -3.60 & 3.56 $\pm$ 0.135 & Oliver et al (1997) \\
 & -2.92 & 1.93 $\pm$ 0.13 & Elbaz et al (2000)\\
 & -3.30 & 2.93 $\pm$ 0.13 & Elbaz et al (2000)\\
 & -3.70 & 3.63 $\pm$ 0.13 & Elbaz et al (2000)\\
 & -4.00 & 3.90 $\pm$ 0.09 & Elbaz et al (2000)\\
60 & 1.00 & -2.12  & Lonsdale et al (1990)\\
& 0.5 & -1.39  & Lonsdale et al (1990)\\
& 0.1 & -0.85  & Lonsdale et al (1990)\\
& -0.22 & -0.38 & Rowan-Robinson et al (1991) \\
& -0.60 & 0.21 $\pm$ 0.1 & Lonsdale et al (1990)\\
 & -0.70 & 0.38 & Lonsdale et al (1990)\\
& -1.00 & 0.76 $\pm$ 0.07 & Hacking and Houck (1987)\\
& -1.30 & 1.46  & Gregorich et al (1995)\\
& -1.30 & 1.30 $\pm$ 0.08 & Hacking and Houck (1987)\\
175 & -1.0 & 1.59 $\pm$ 0.1 & Kawara et al (1998),Guiderdoni et al (1998)\\
850 & -2.40 & 2.95 +0.15,-0.25 & Hughes et al (1998)\\
  & -2.70 & 3.40 +0.15,-0.25 & Hughes et al (1998)\\
  & -2.55 & 3.25 +0.13,-0.17 & Eales et al (1998)\\
  & -2.40 & 3.39 $\pm$ 0.19 & Smail et al (1997)\\
  & -2.65 & 3.37 $\pm$ 0.30 & Barger et al (1999)\\
  & -3.00 & 3.90 +0.14,-0.21 & Blain et al (1999c)\\
  & -2.155 & 2.56 +0.23,-0.52  & Fox (2000)\\
\end{tabular}
\end{table}

\clearpage

\begin{table}
\caption{Observed submillimetre background}
\begin{tabular}{crrr}
& & \\
wavelength ($\mu$m) & $log_{10}  \nu I_{\nu} (nW/m^{2}/sr)$ & reference\\
& & \\
140 & 1.51 +0.14, -0.23 & Schlegel et al (1998)\\
 & 1.40 +0.11, -0.15 & Hauser et al (1998)\\
 & 1.10 $\pm$ 0.10 & Fixsen et al (1998)\\
 & 1.18 +0.15, -0.23 & Lagache et al (1999)\\
240 & 1.00 $\pm$ 0.20 & Puget et al (1996)\\
 & 1.23 +0.09, -0.12 & Schlegel et al (1998)\\
 & 1.15 +0.08, -0.11 & Hauser et al (1998)\\
 & 1.04 $\pm$ 0.10 & Fixsen et al (1998)\\
 & 1.06 +0.06, -0.08 & Lagache et al (1999)\\
350 & 0.85 $\pm$ 0.20 & Puget et al (1996)\\
 & 0.78 $\pm$ 0.10 & Fixsen et al (1998)\\
 & 0.85 +0.13, -0.21 & Lagache et al (1999)\\
500 & 0.37 & Fixsen et al (1998)\\
 & 0.59 +0.16, -0.26 & Lagache et al (1999)\\
750 & -0.12 & Fixsen et al (1998)\\
 & -0.05 +0.11, -0.14 & Lagache et al (1999)\\
850 & 0.00 $\pm$ 0.15 & Puget et al (1996)\\
 & -0.30 $\pm$ 0.10 & Fixsen et al (1998)\\
\end{tabular}
\end{table}

\begin{table}
\caption{Best fit star-formation history models for $\Omega_o = 1/\Omega_o=0.3/\Lambda=0.7$}
\begin{tabular}{crrrrrr}
 & & & & & \\
cosmological & P & Q & $l_{60}$ & $\xi$ & $\Omega_*$ \\
model & & & $h L_{odot} Mpc^{-3}$ & & \\
 & & & & & \\
$\Omega_o$ = 1 & 1.2 & 5.4 & 4.3 & 5.70 & 0.0027 $h^{-1}$ \\
$\Omega_o$ = 0.3 & 2.1 & 7.3 & 4.4 & 6.66 & 0.0032 $h^{-1}$ \\
$\Lambda$ = 0.7 & 3.0 & 9.0 & 4.1 & 7.25 & 0.0033 $h^{-1}$ \\
\end{tabular}
\end{table}

\begin{table}
\caption{Predicted integrated background flux from different components, for 
$\Omega_o = 1/\Omega_o=0.3/\Lambda=0.7$}
\begin{tabular}{crrrrr}
 & & & & \\
component & $\nu I_{\nu}$ at 0.1-5 $\mu$m & 5-100 $\mu$m & 100-1600 $\mu$m & \bf {total} \\
& $(nW/m^{2}/sr)$  & & & \\
 & & & & \\
 cirrus+starlight & 13.81/15.73/18.07 & 5.30/6.94/9.01 & 10.09/14.30/18.45 & 28.20/36.97/45.36\\
 M82 starburst & 1.25/1.71/2.11 & 2.68/3.56/4.35 & 1.83/2.63/3.25 & 5.76/7.90/9.71\\
 A220 starburst & 0.0004/0.0006/0.0007 & 0.004/0.005/0.006 & 0.0074/0.0103/0.0122 & 0.0118/0.0159/0.0189\\
 AGN + dust torus & 0.243/0.320/0.432 & 0.247/0.333/0.431 & 0.0074/0.0106/0.0116 & 0.490/0.674/0.875\\
 & & & & \\
 \bf{total(all types)} & \bf{14.34/17.75/20.61} & \bf{8.21/10.86/13.63} & \bf{11.91/16.95/21.73} & \bf{34.46/45.56/55.97}\\
\end{tabular}
\end{table}

\clearpage

\begin{table}
\caption{Confusion limit for different infrared and submillimetre telescopes, for $\Omega_o$ =1 model}
\begin{tabular}{crrrrrrrrrrr}
 & & & & & & & & & & &\\
Telescope & D(m) & $\lambda (\mu m)$ & 2 $\theta_{conf}$ & $N_{conf}$  & 
lg $S_{conf}$ & redshift & distribn & (fraction) & & &  \\
& & & & ($/$sq dg) & (Jy) & 0-1 & 1-2 & 2-3 & 3-4 & 4-5 & $>$5  \\
 & & & & & & & & & & &\\
 ISO & 0.6 & 15 & 6.2" & 10800 & -4.22 & 0.7011 & 0.2858 & 0.0077 & 0.0009 & 0.0001 & 0.0\\
     &     & 90 & 37.1" & 300 & -1.66 & 0.9708 & 0.0290 & 0.0002 & 0.0 & 0.0 & 0.0\\
     & & 175 & 72.2" & 79 & -0.97 & 0.9785 & 0.0212 & 0.0002 & 0.0 & 0.0 & 0.0\\
     & & & & & & & & & & &\\
 SCUBA & 15. & 450 & 7.4" & 590 & -1.90 &  0.3287 & 0.4850 & 0.1583 & 0.0258 & 0.0020 & 0.0002\\
 & & 850 & 14.0" & 2100 & -2.63 & 0.1082 & 0.3193 & 0.2801 & 0.1772 & 0.0746 & 0.0406 \\
 & & & & & & & & & & &\\
 SIRTF & 0.85 & 8.0 & 2.33" & 75780 & -6.17 & 0.3829 & 0.4525 & 0.0868 & 0.0413 & 0.0172 & 0.0191 \\
 & & 25 & 7.3" & 7760 & -3.87 & 0.6062 & 0.3484 & 0.0428 & 0.0024 & 0.0001 & 0.0  \\
 & & 70 & 20.4" & 990 & -2.33 &  0.9314 & 0.0675 & 0.0010 & 0.0 & 0.0 & 0.0 \\
 & & 160 & 46.6" & 190 & -1.23 &  0.9741 & 0.0253 & 0.0003 & 0.0 & 0.0 & 0.0 \\
 & & & & & & & & & & &\\
 ASTRO-F & 0.6 & 60 & 25" & 700 & -2.36 &  0.9334 & 0.0655 & 0.0010 & 0.0 & 0.0 & 0.0 \\
 & & 175 & 72.2" & 79 & -0.97 &  0.9785 & 0.0212 & 0.0002 & 0.0 & 0.0 & 0.0 \\
 & & & & & & & & & & & \\
 FIRST & 3.5 & 90 & 6.4" & 10185 & -3.13 & 0.7456 & 0.2316 & 0.0211 & 0.0016 & 0.0001 & 0.0 \\ 
 & & 120 & 8.5" & 5730 & -2.50 &  0.8366 & 0.1546 & 0.0083 & 0.0004 & 0.0 & 0.0 \\
 & & 175 & 12.4" & 2680 & -1.96 &  0.8261 & 0.1683 & 0.0052 & 0.0003 & 0.0 & 0.0 \\
 & & 250 & 17.7" & 1320 &  -1.73 & 0.7302 & 0.2609 & 0.0085 & 0.0004 & 0.0 & 0.0 \\
 & & 350 & 24.8" & 670 & -1.70 &  0.5469 & 0.4077 & 0.0432 & 0.0021 & 0.0001 & 0.0 \\
 & & 500 & 35.4" & 330 & -1.78 &  0.3454 & 0.4838 & 0.1469 & 0.0223 & 0.0016 & 0.0001 \\
 & & 600 & 42.5" & 230 & -1.88 &  0.2426 & 0.4805 & 0.2145 & 0.0544 & 0.0071 & 0.0010 \\
 & & & & & & & & & & & \\
 PLANCK & (1.5) & 350 & 5' & 4.6 & -0.83 & 0.8104 & 0.1851 & 0.0043 & 0.0001 & 0.0 & 0.0 \\
 & & 550 & 5' & 4.6 & -1.13 &  0.4454 & 0.4220 & 0.1222 & 0.0097 & 0.0005 & 0.0001 \\
 & & 850 & 5' & 4.6 & -1.63 &  0.1555 & 0.3732 & 0.3086 & 0.1310 & 0.0276 & 0.0040 \\
\end{tabular}
\end{table}







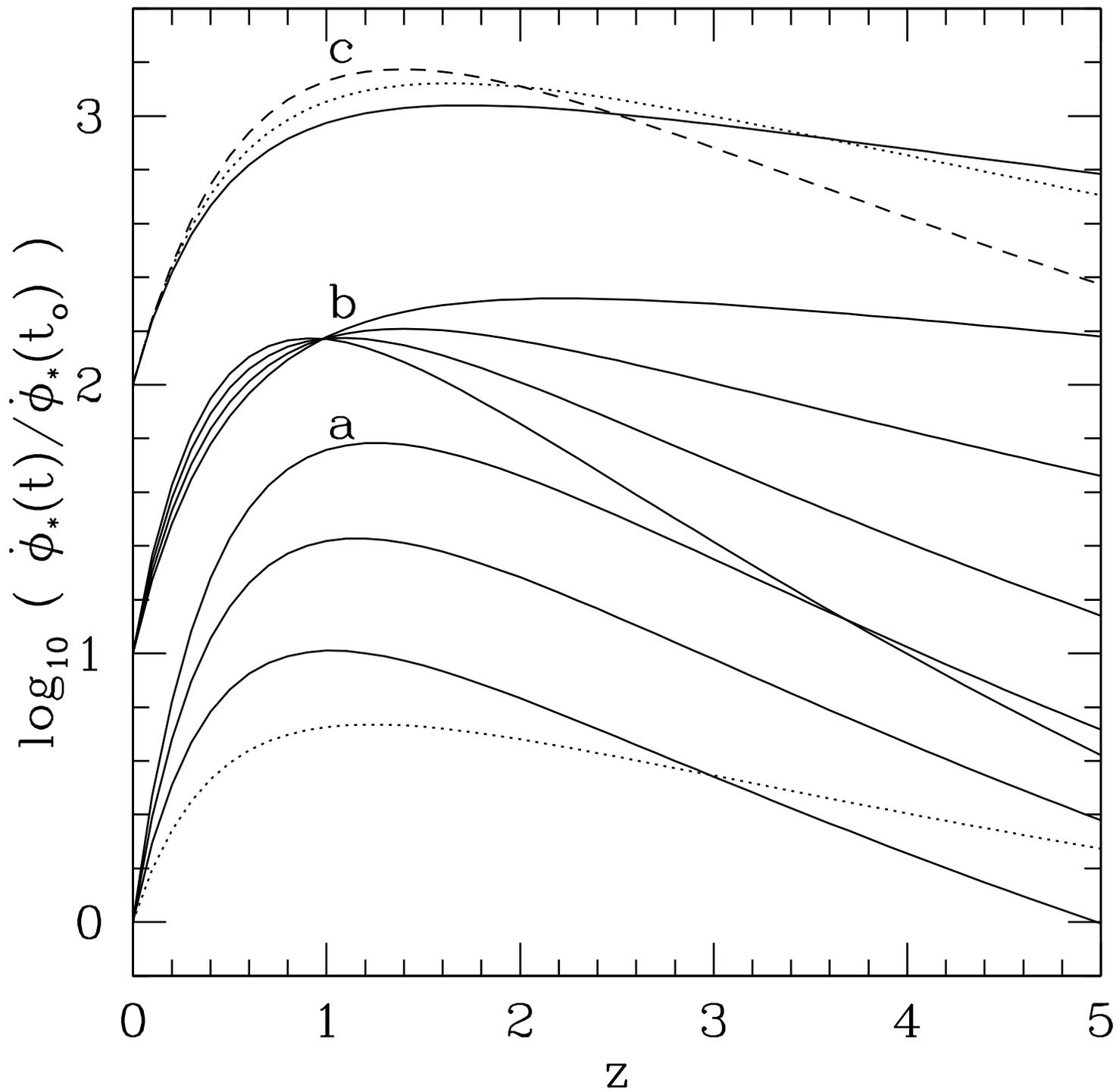
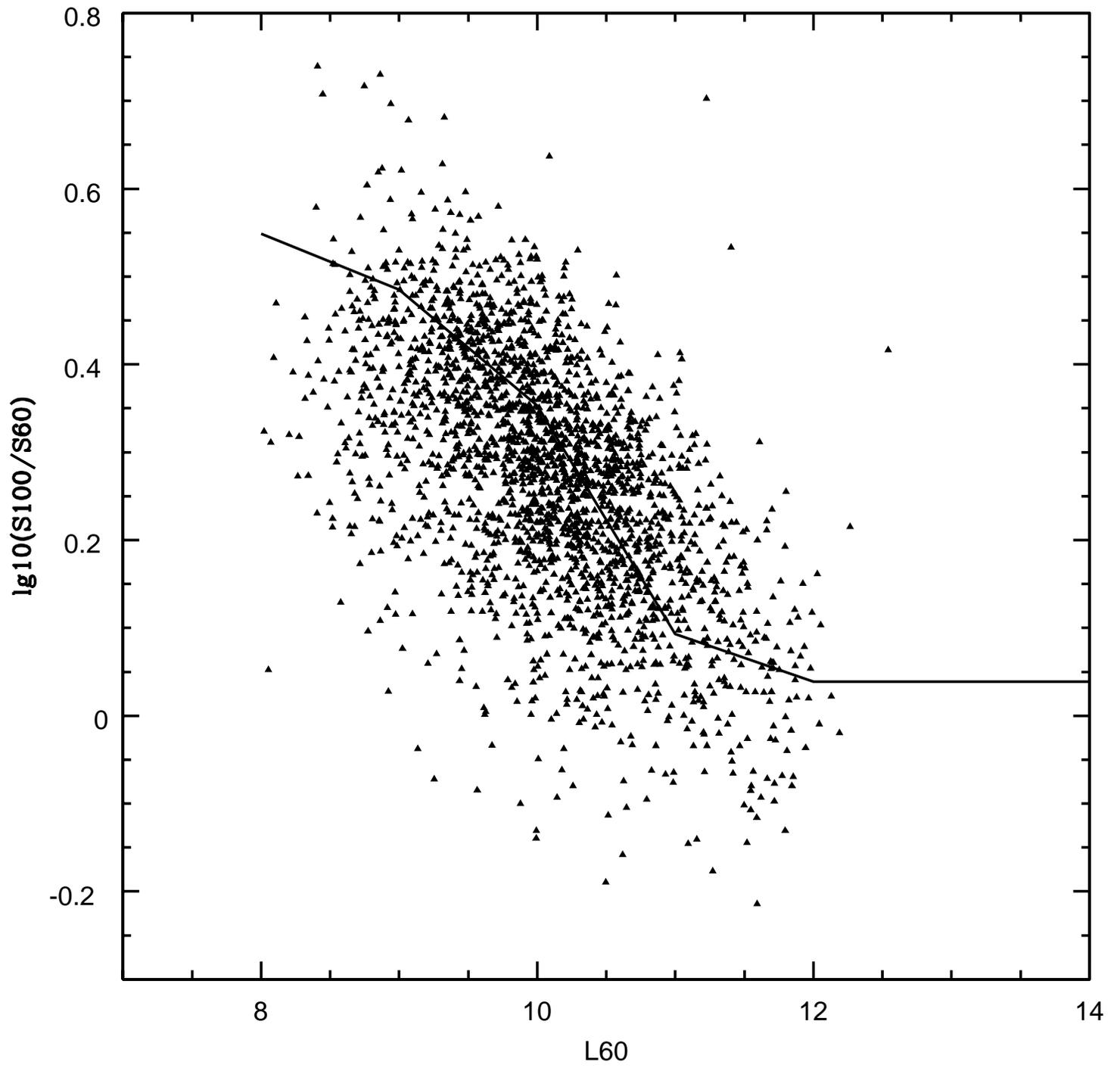
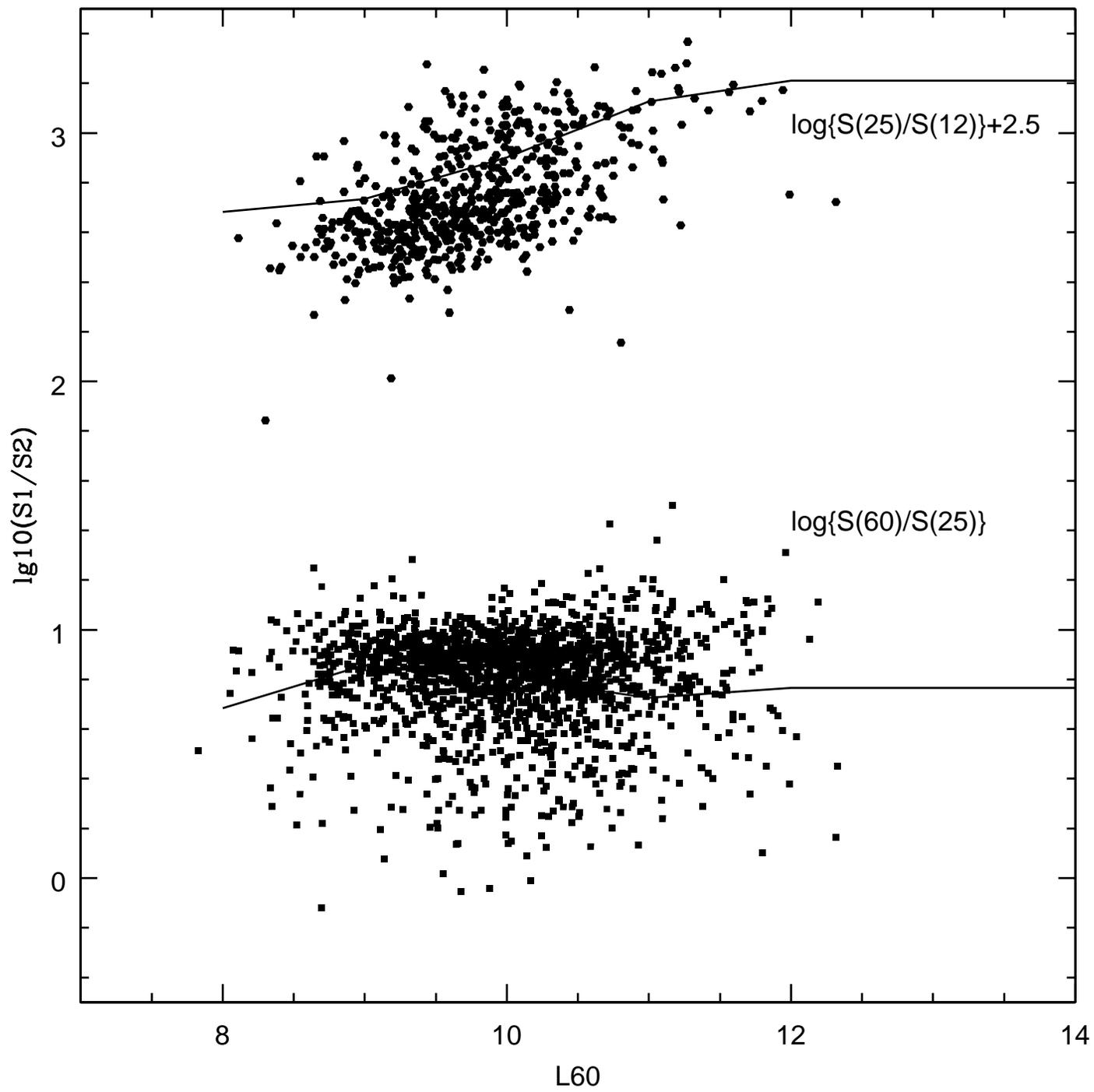
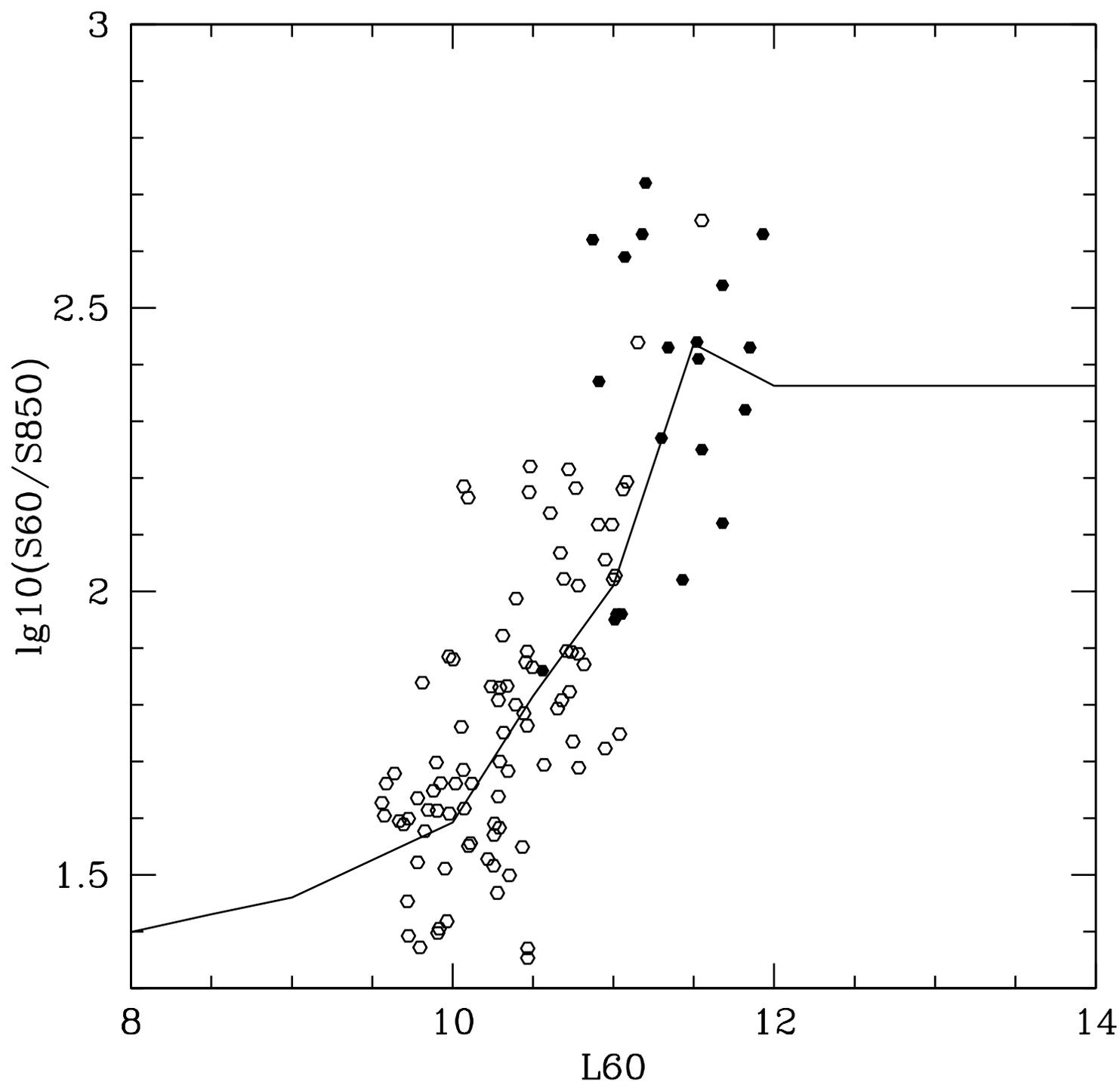
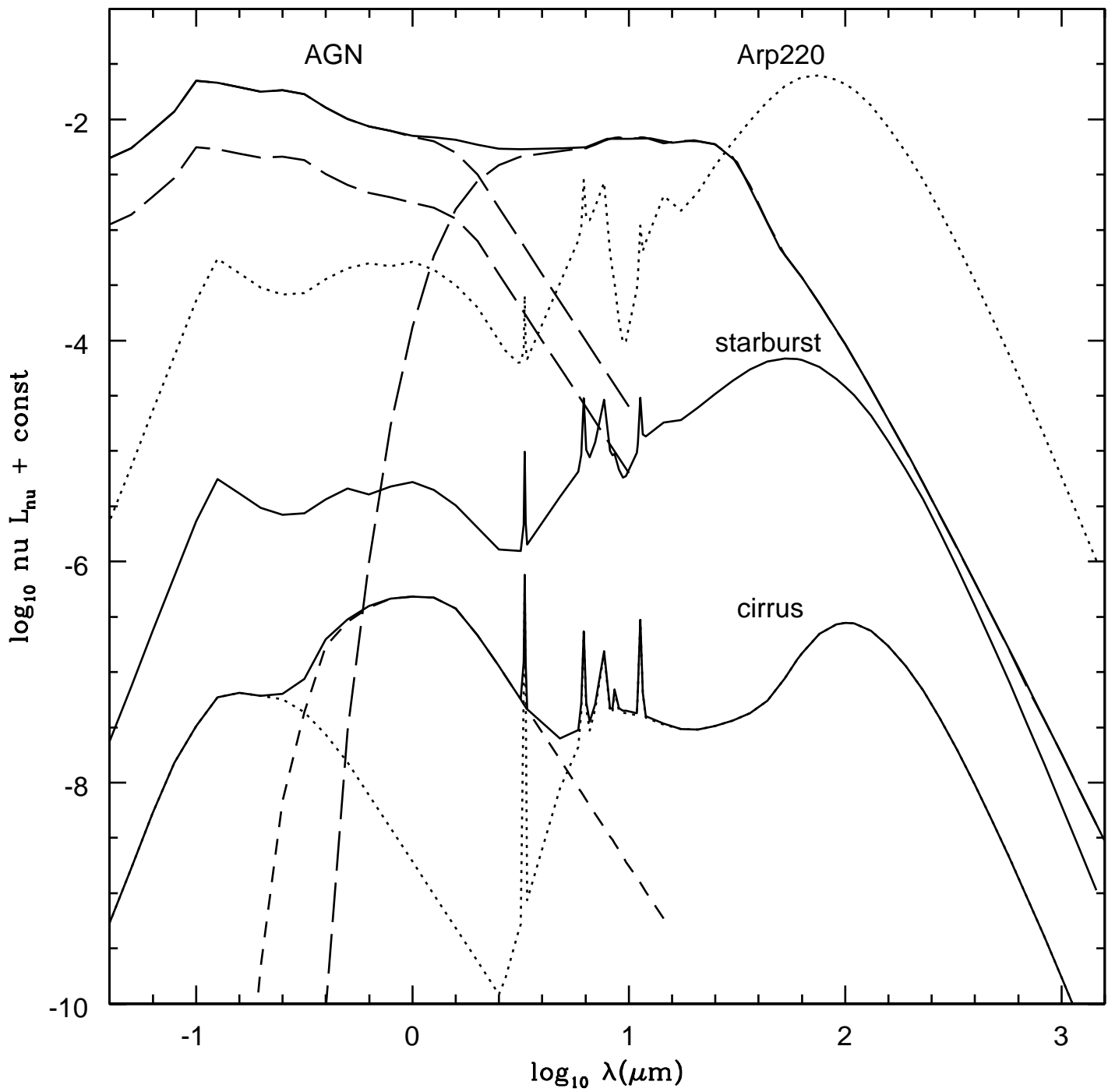
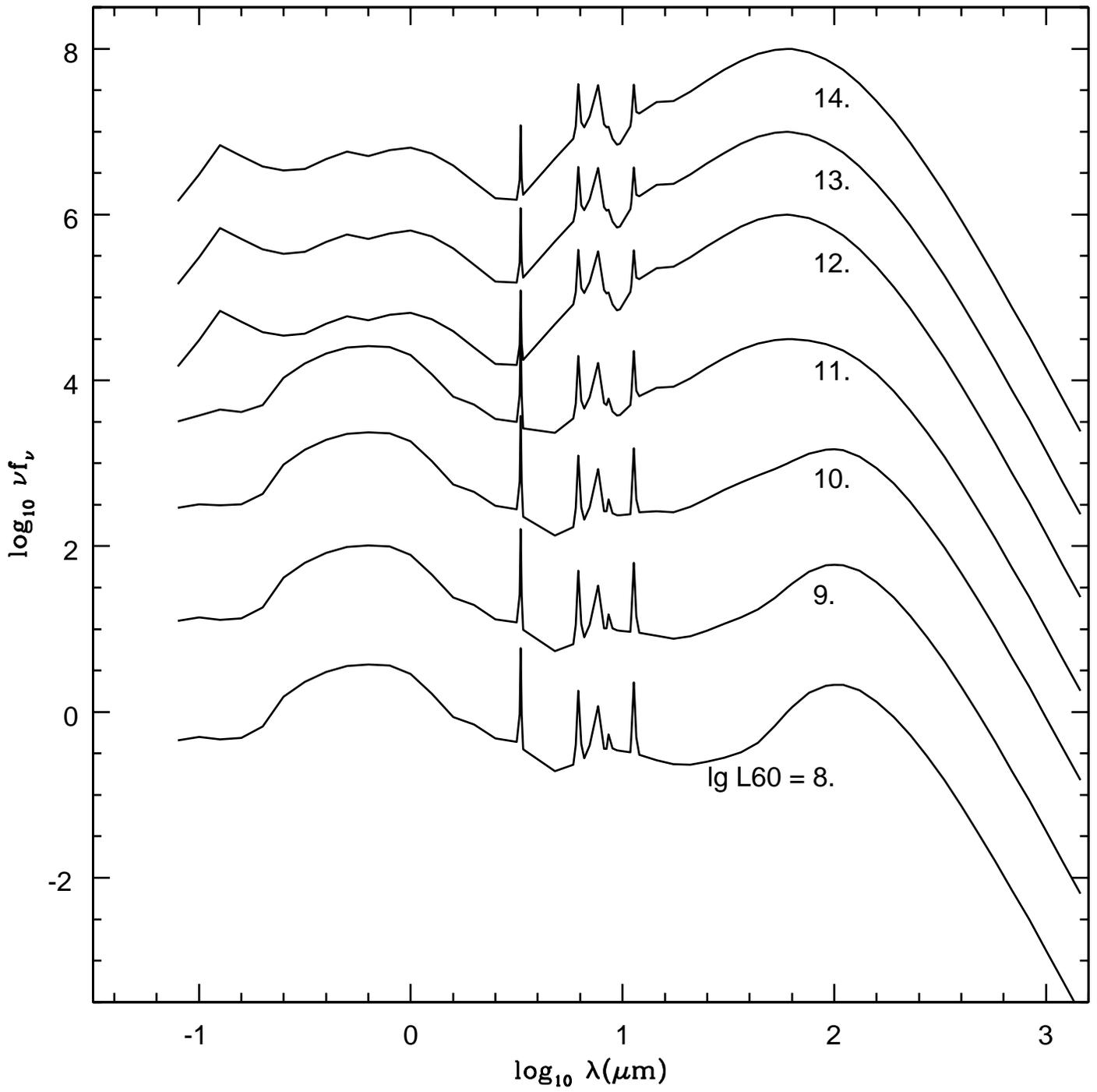
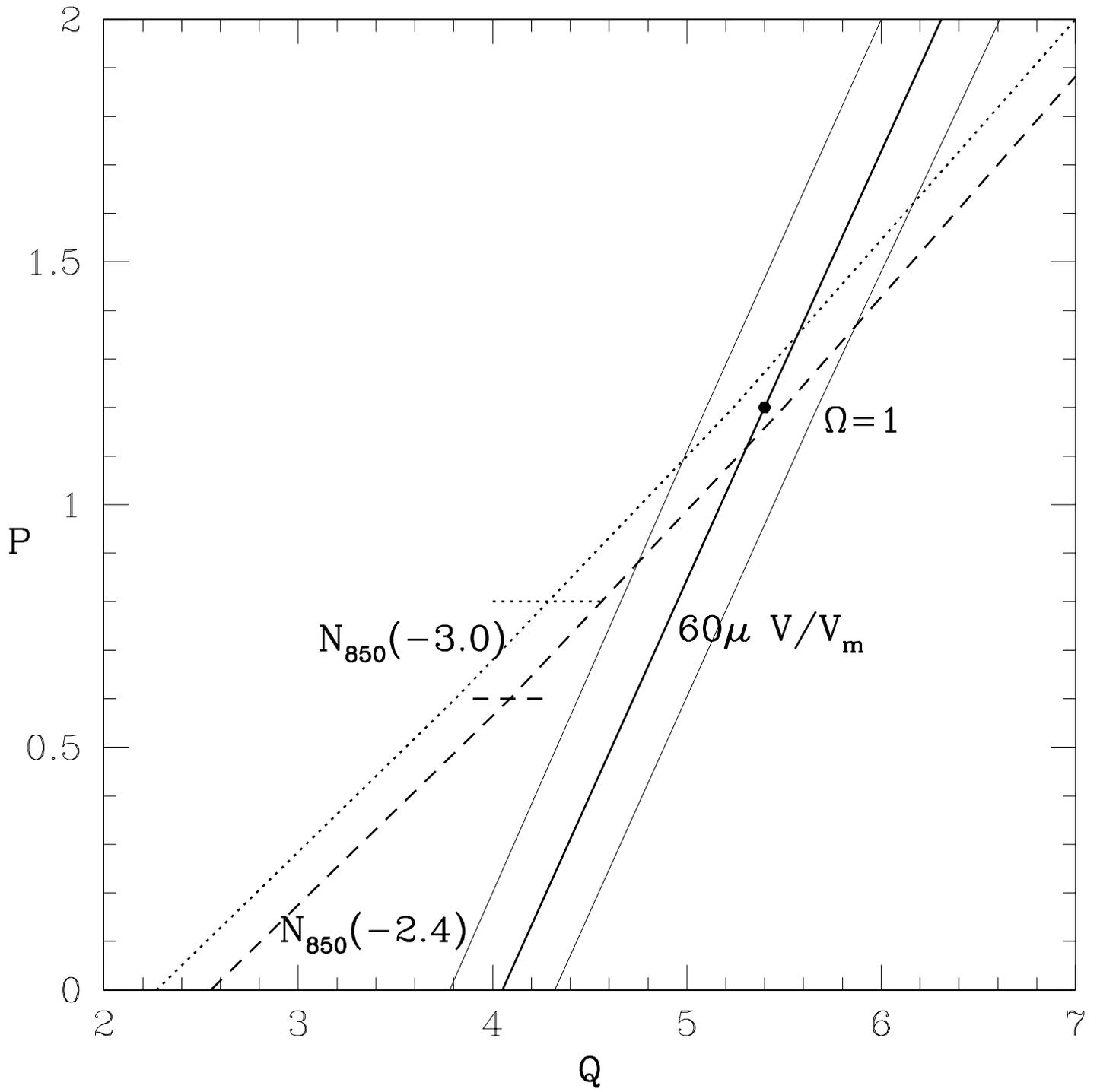
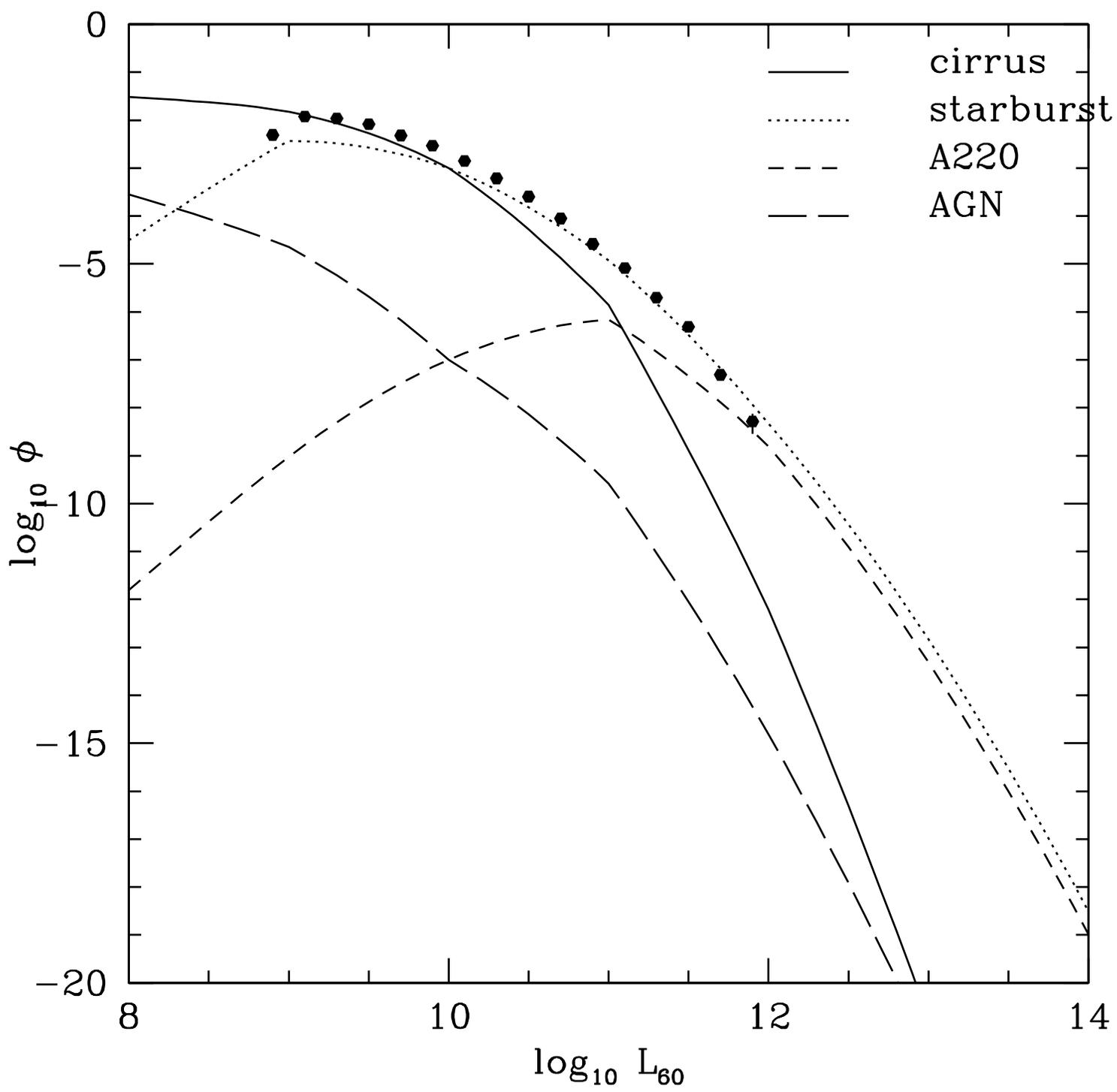
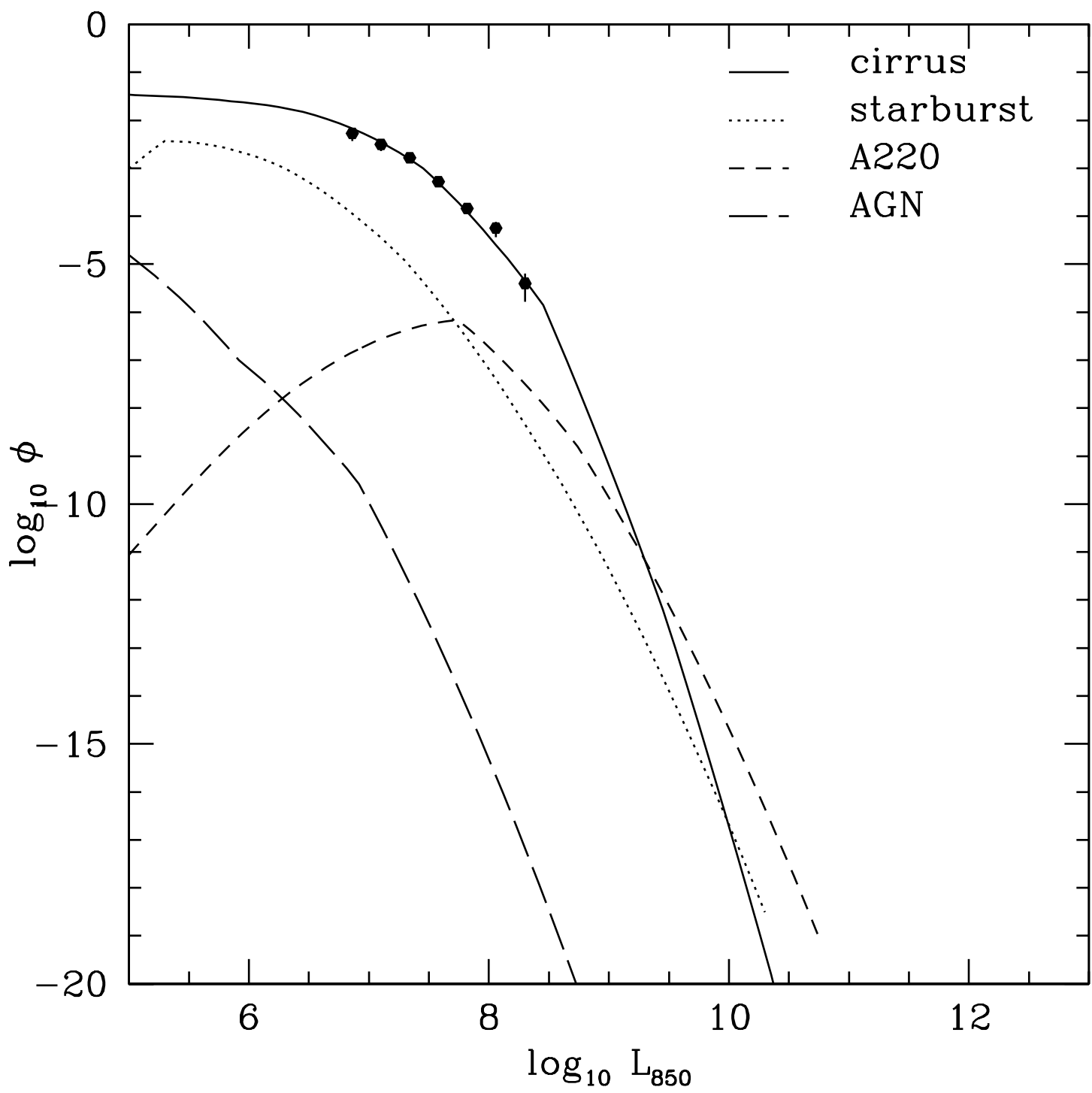
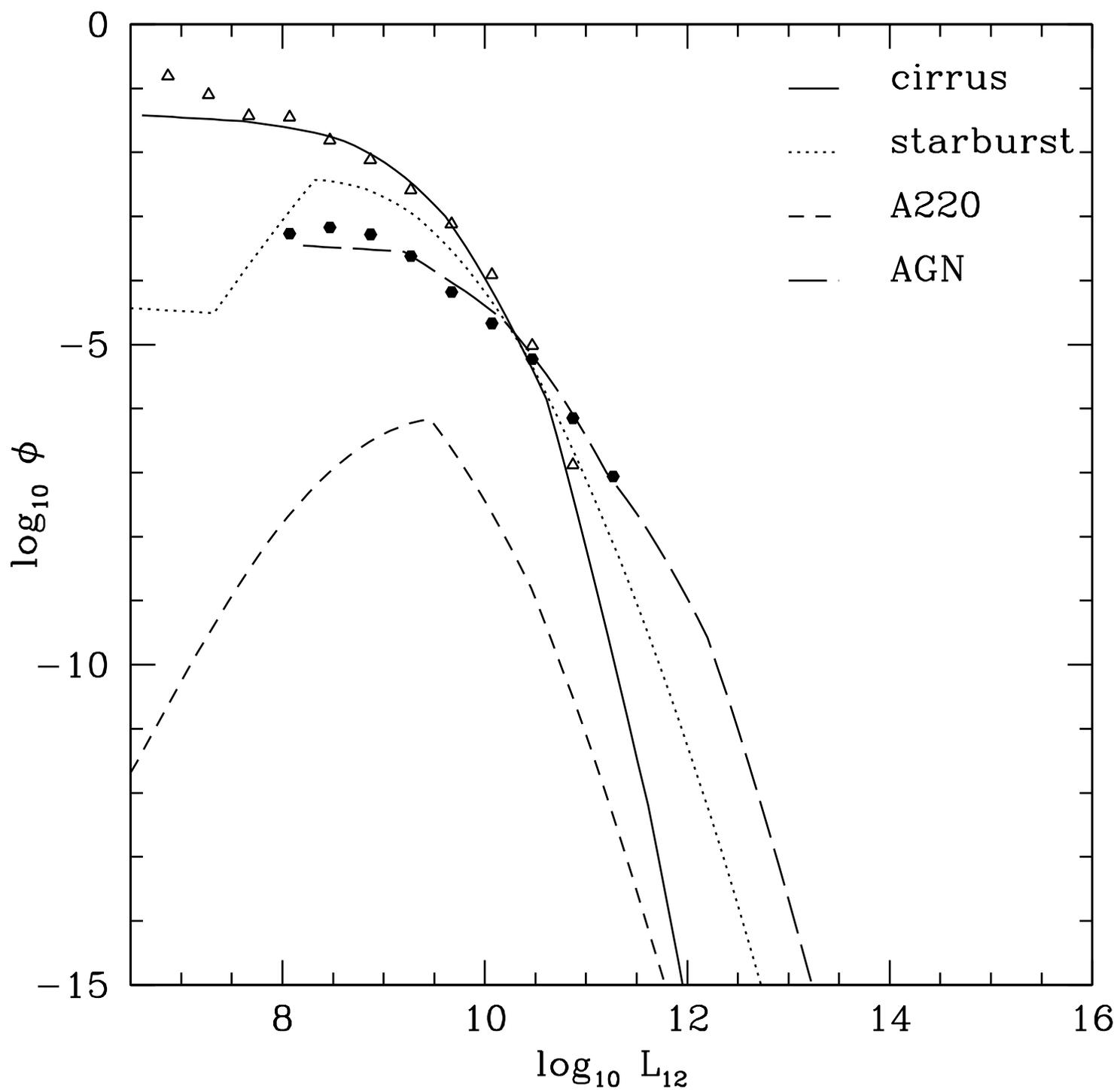
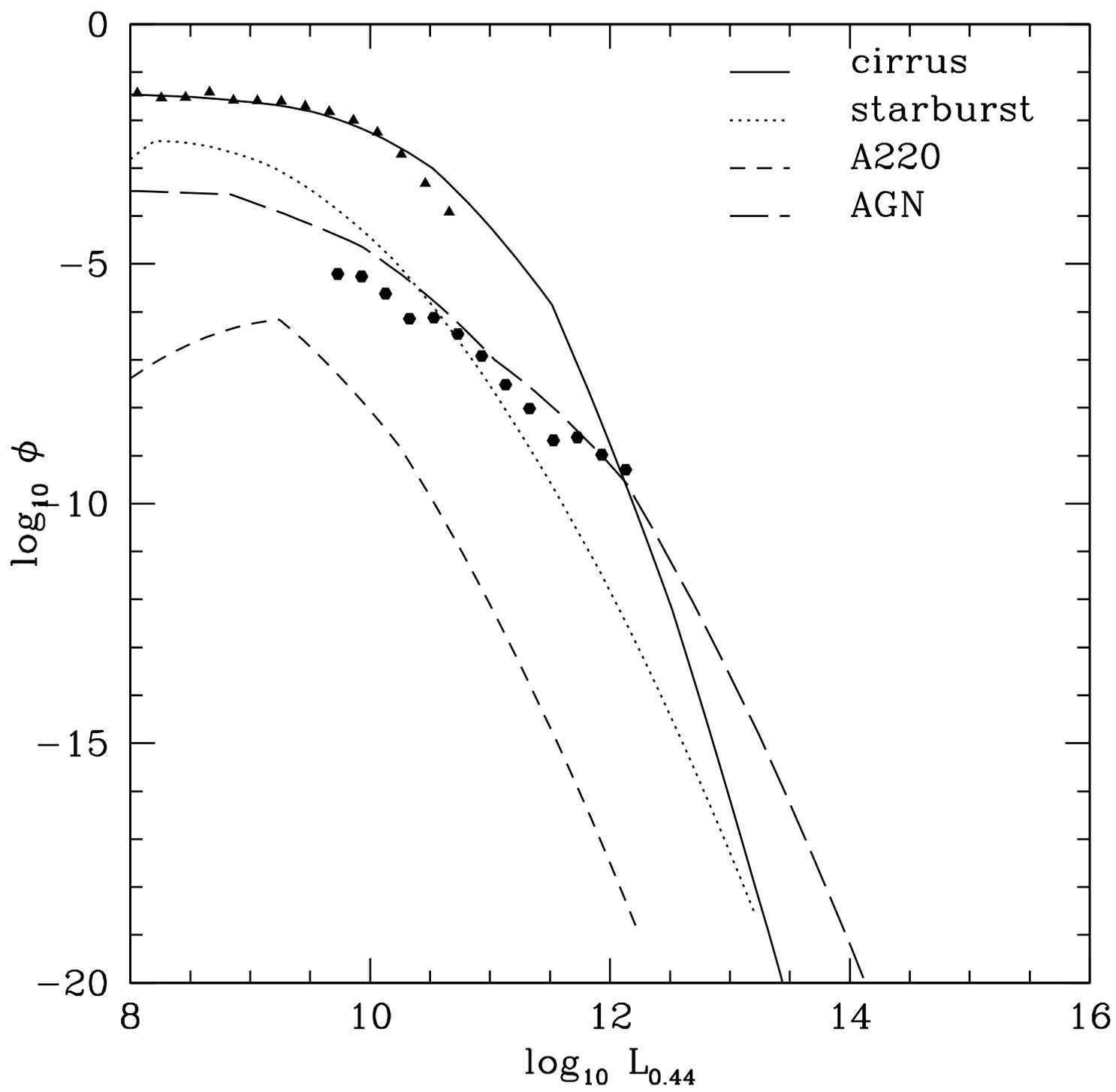
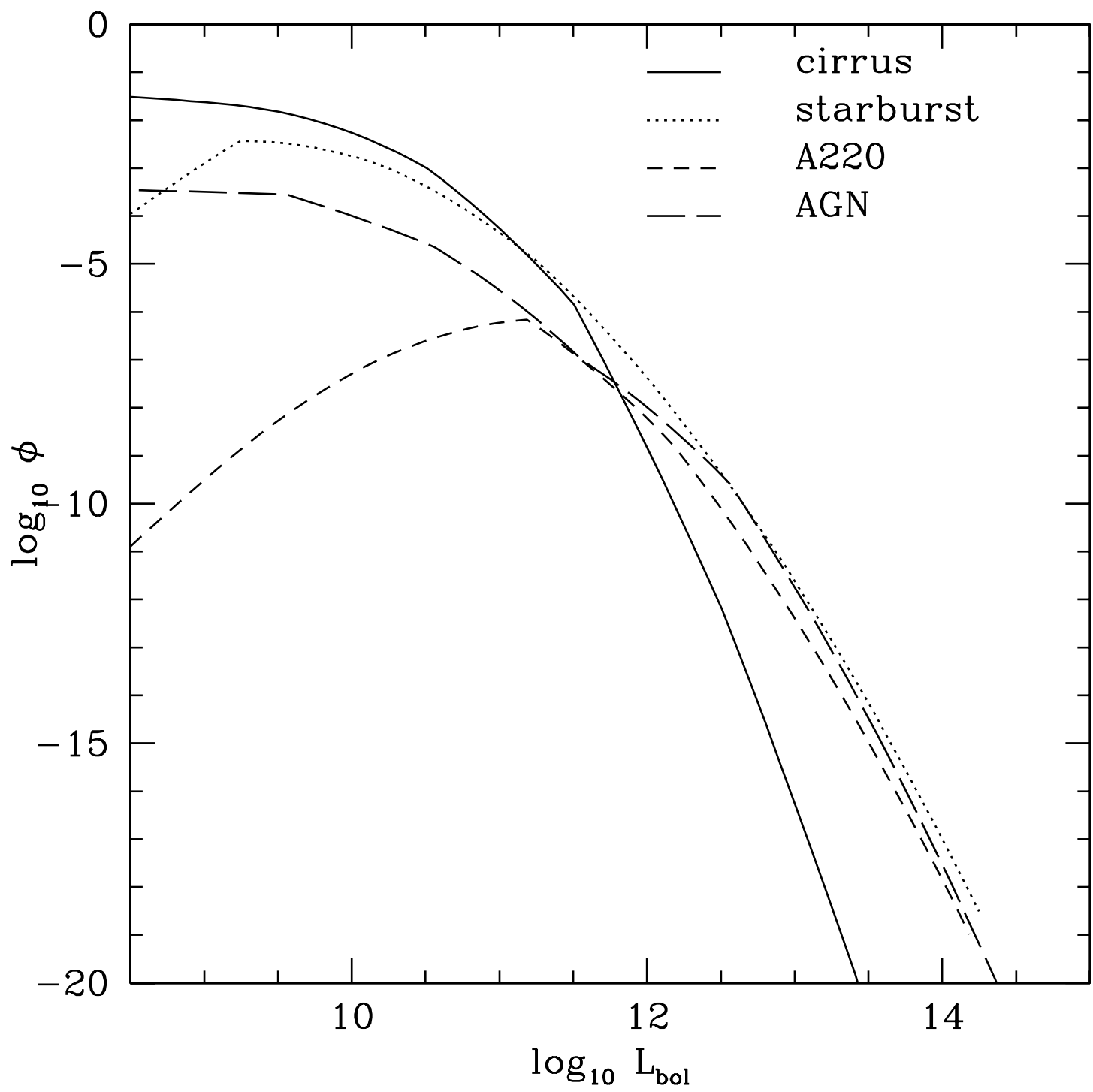
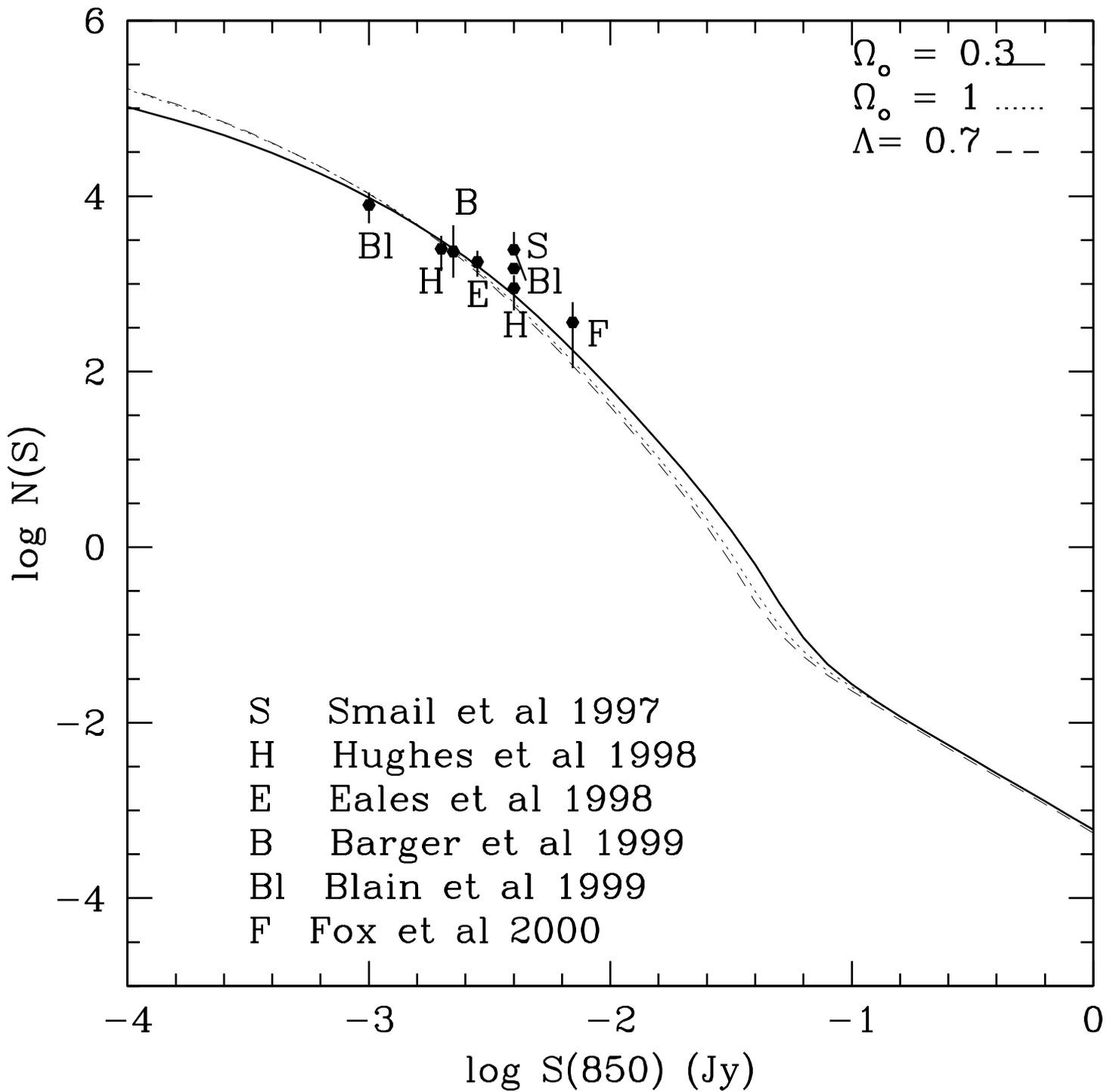
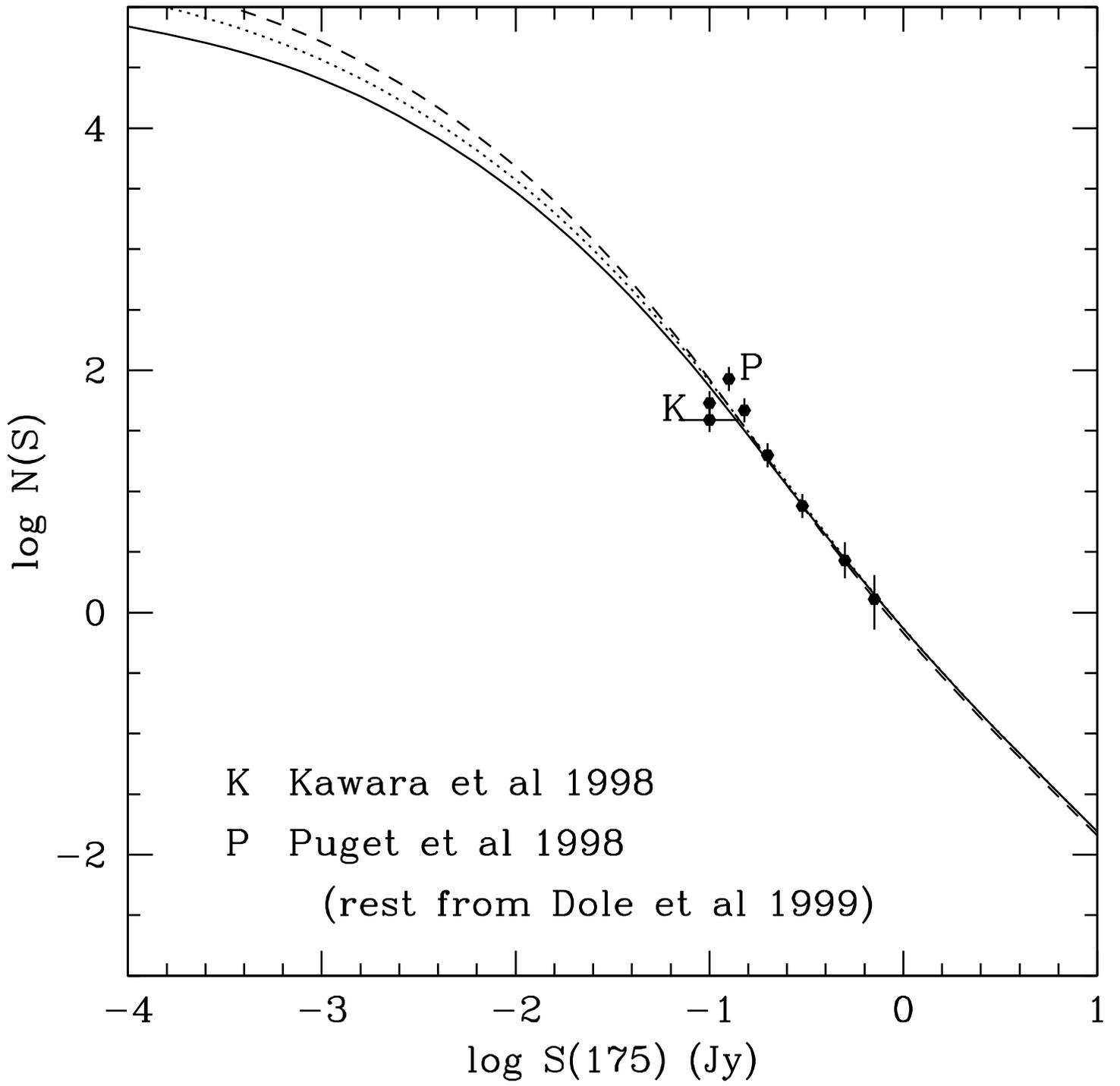
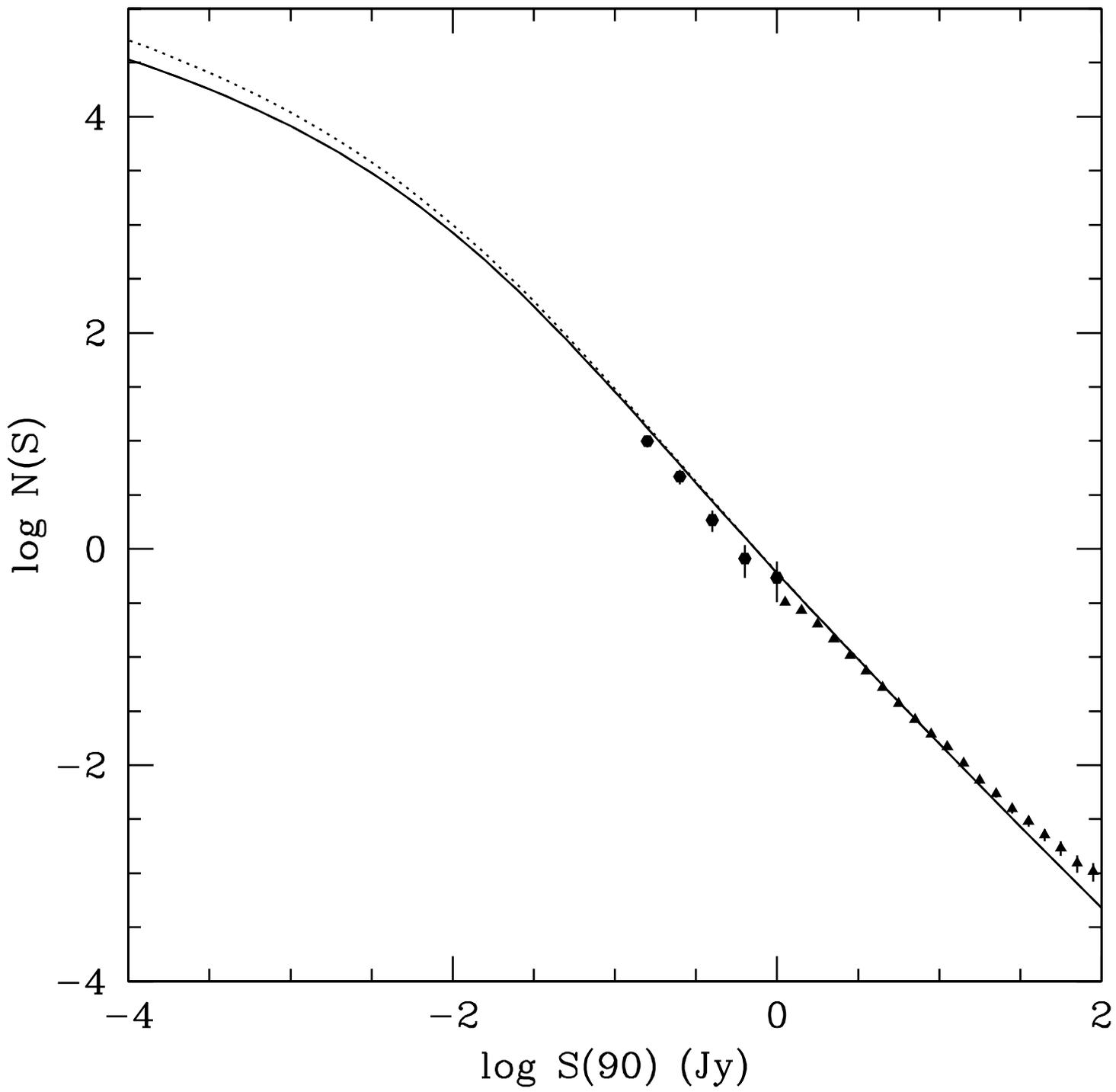
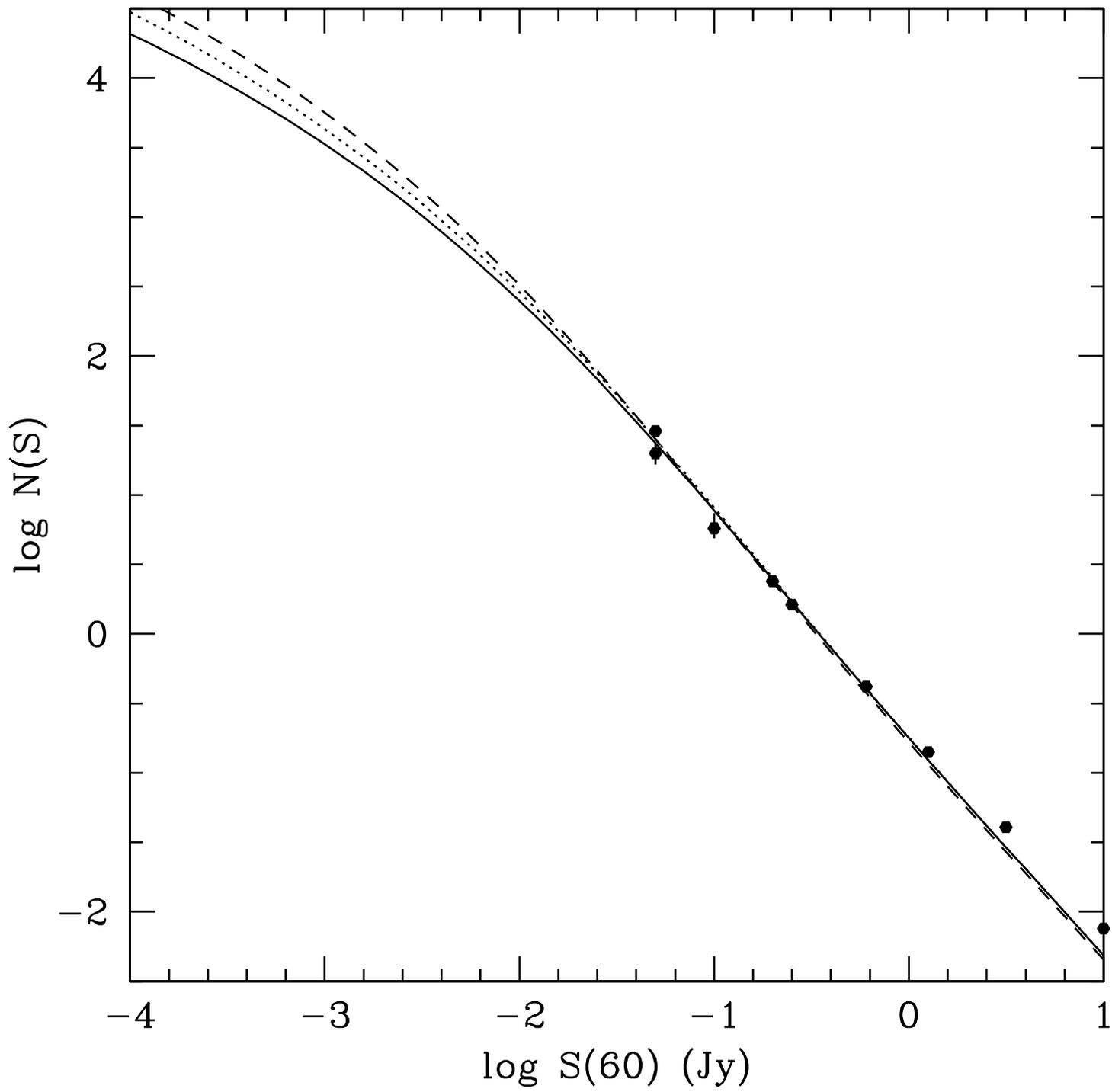
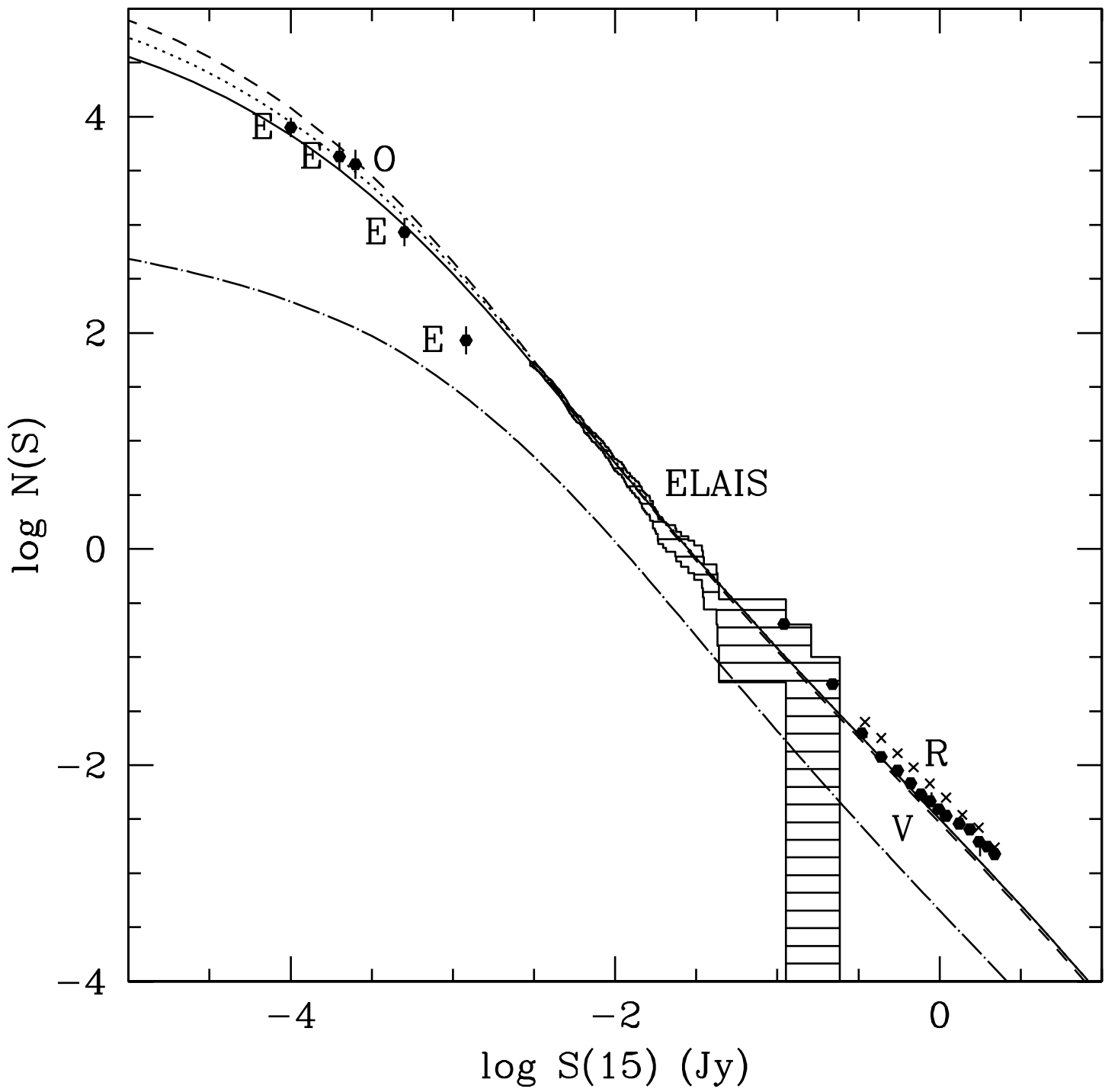
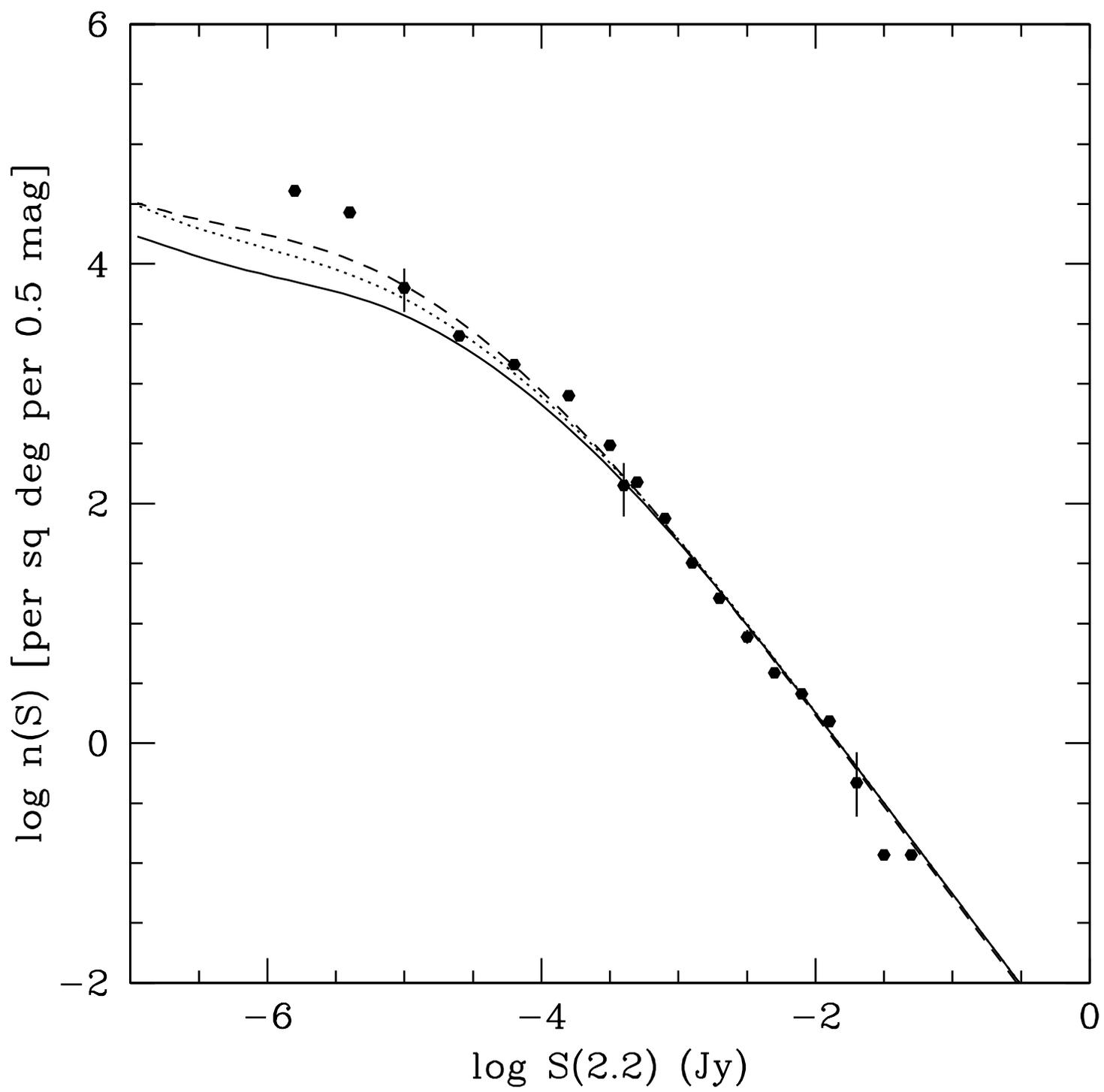
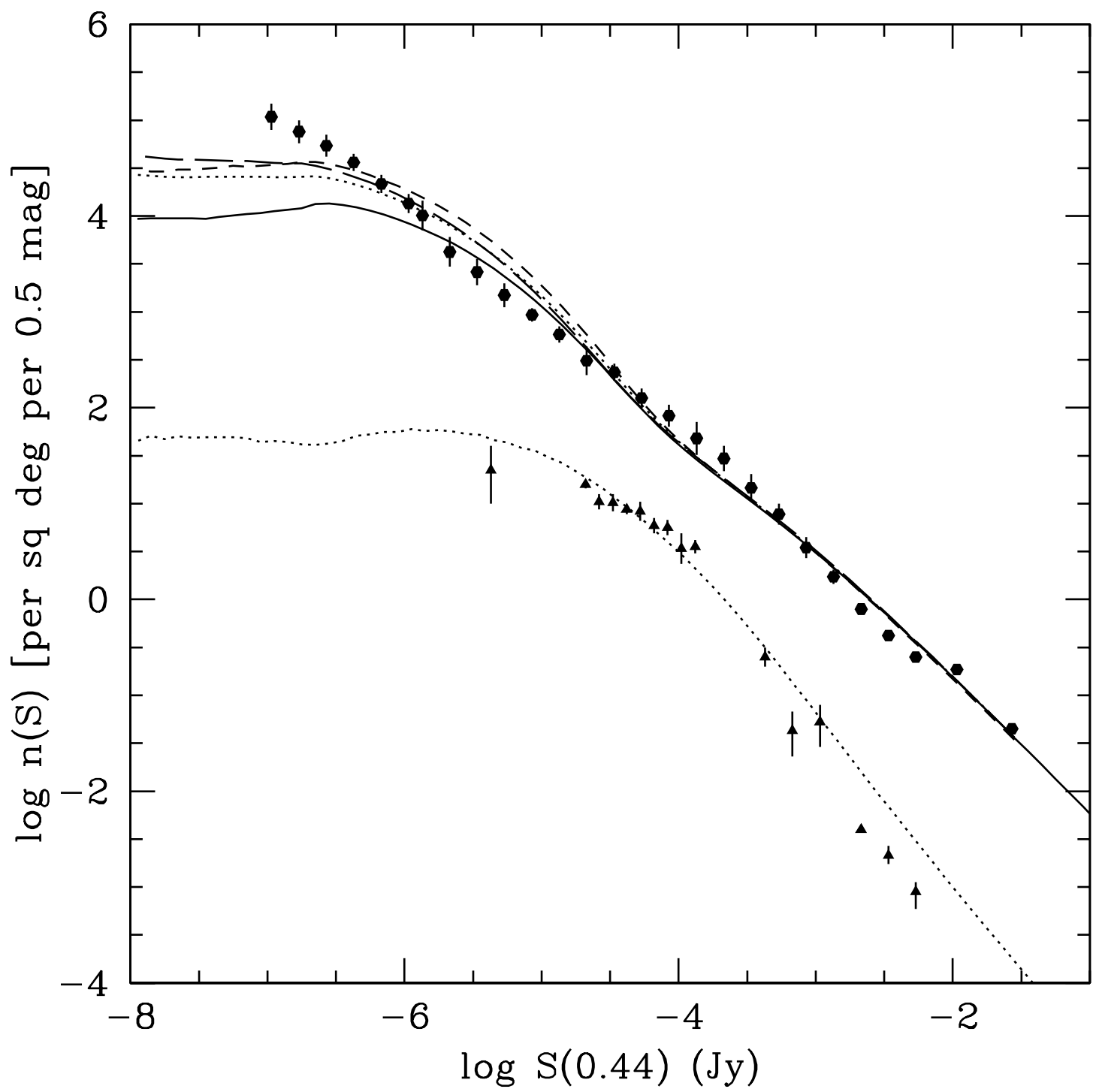
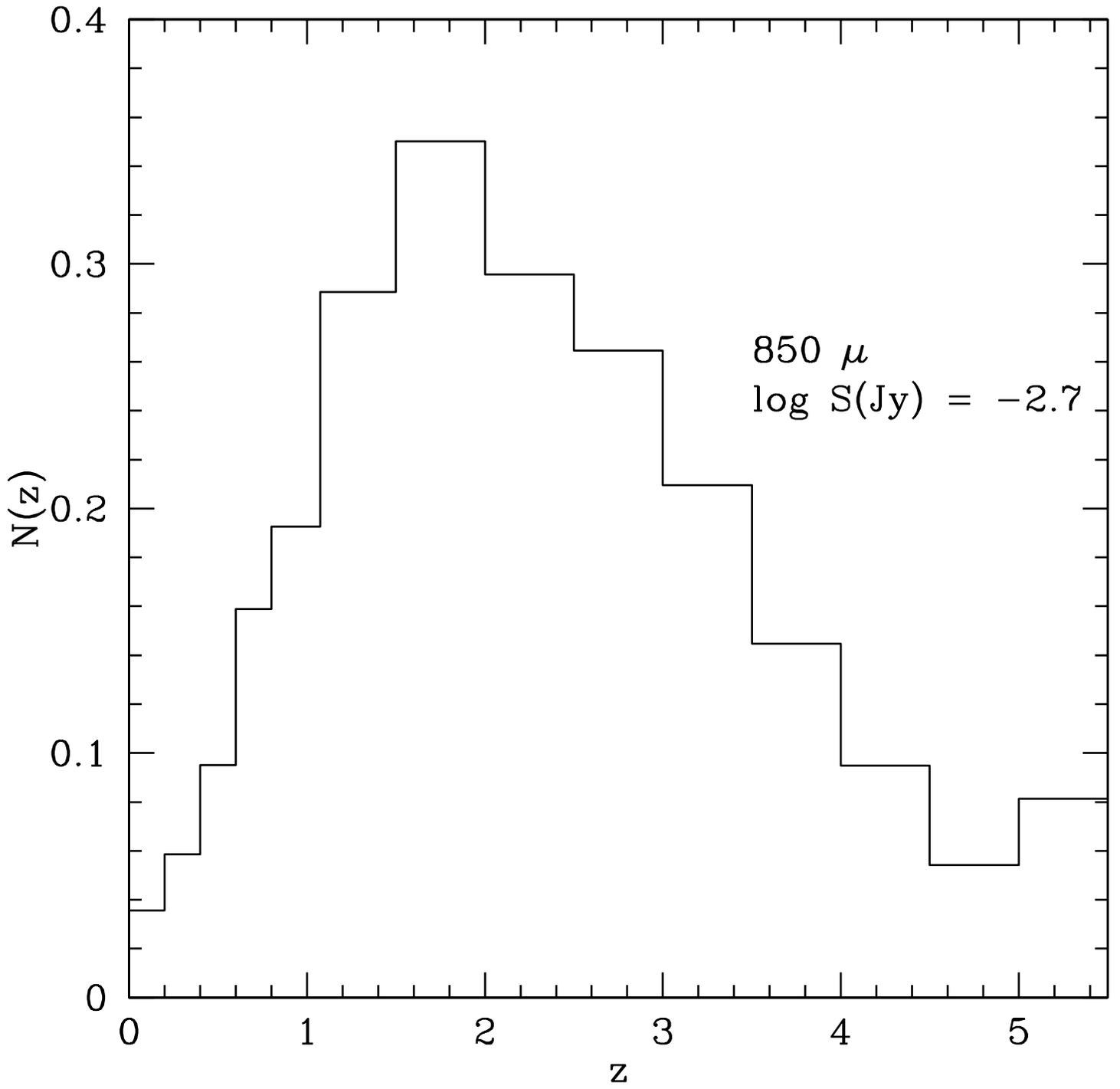
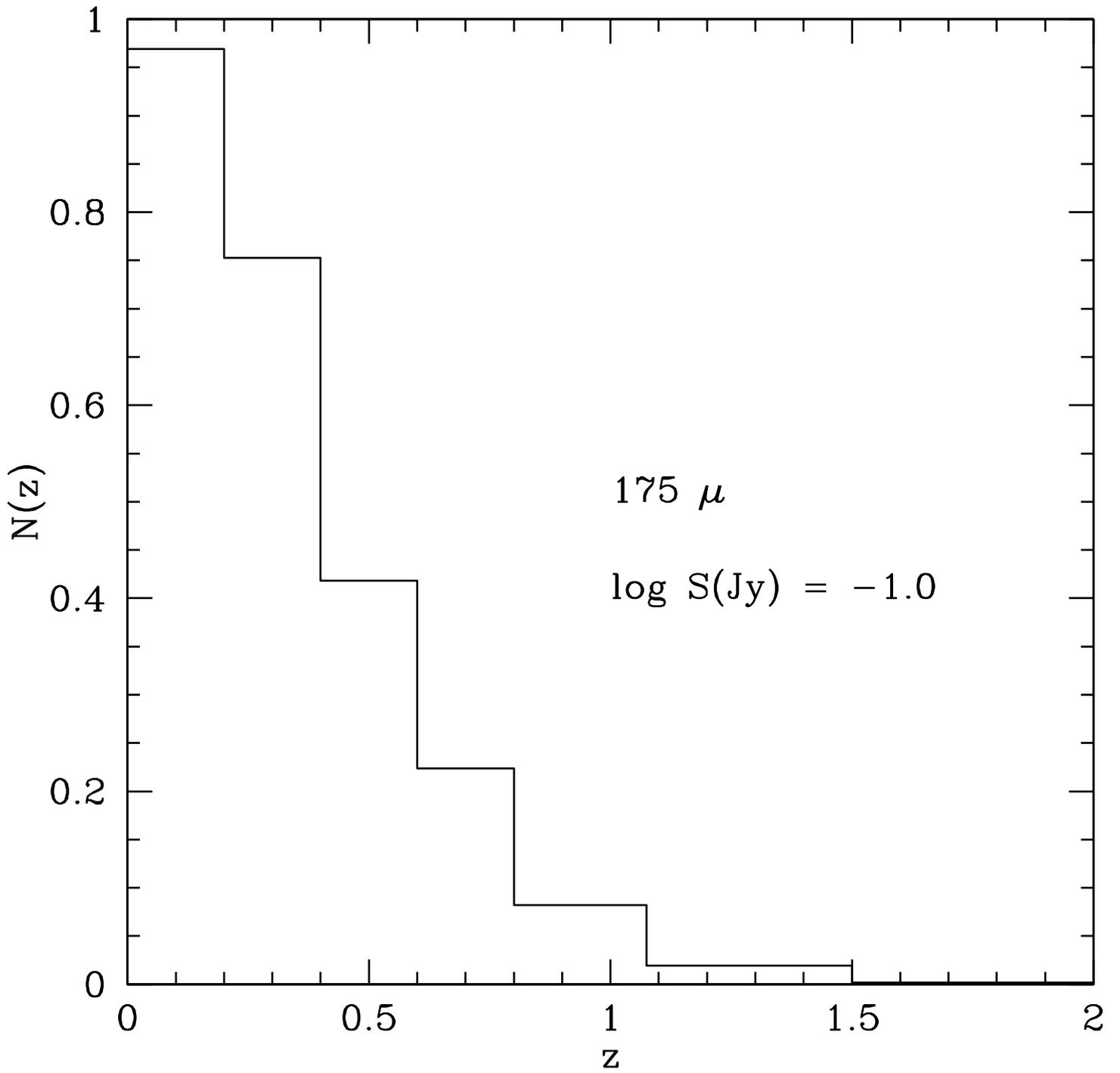
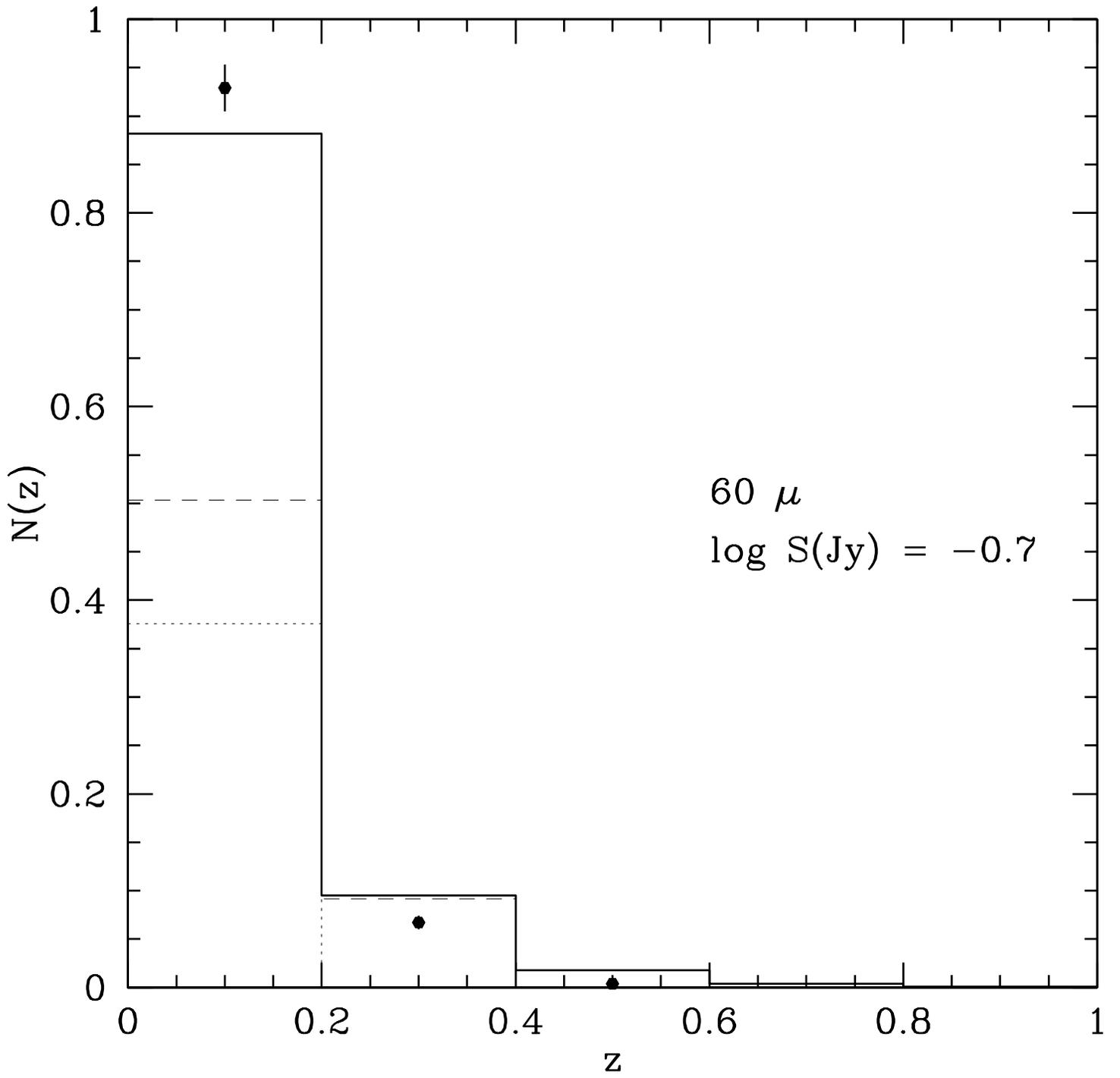
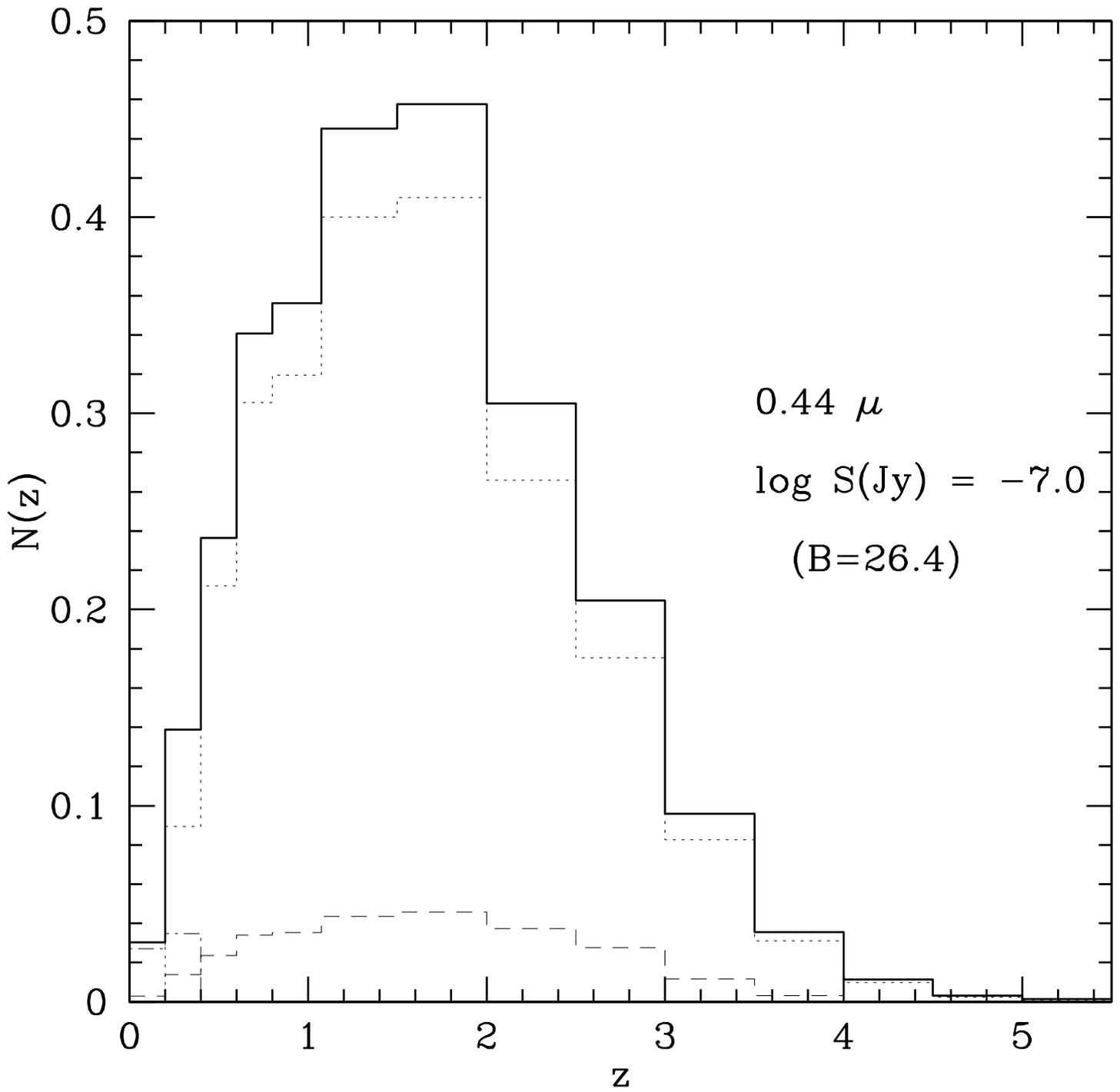
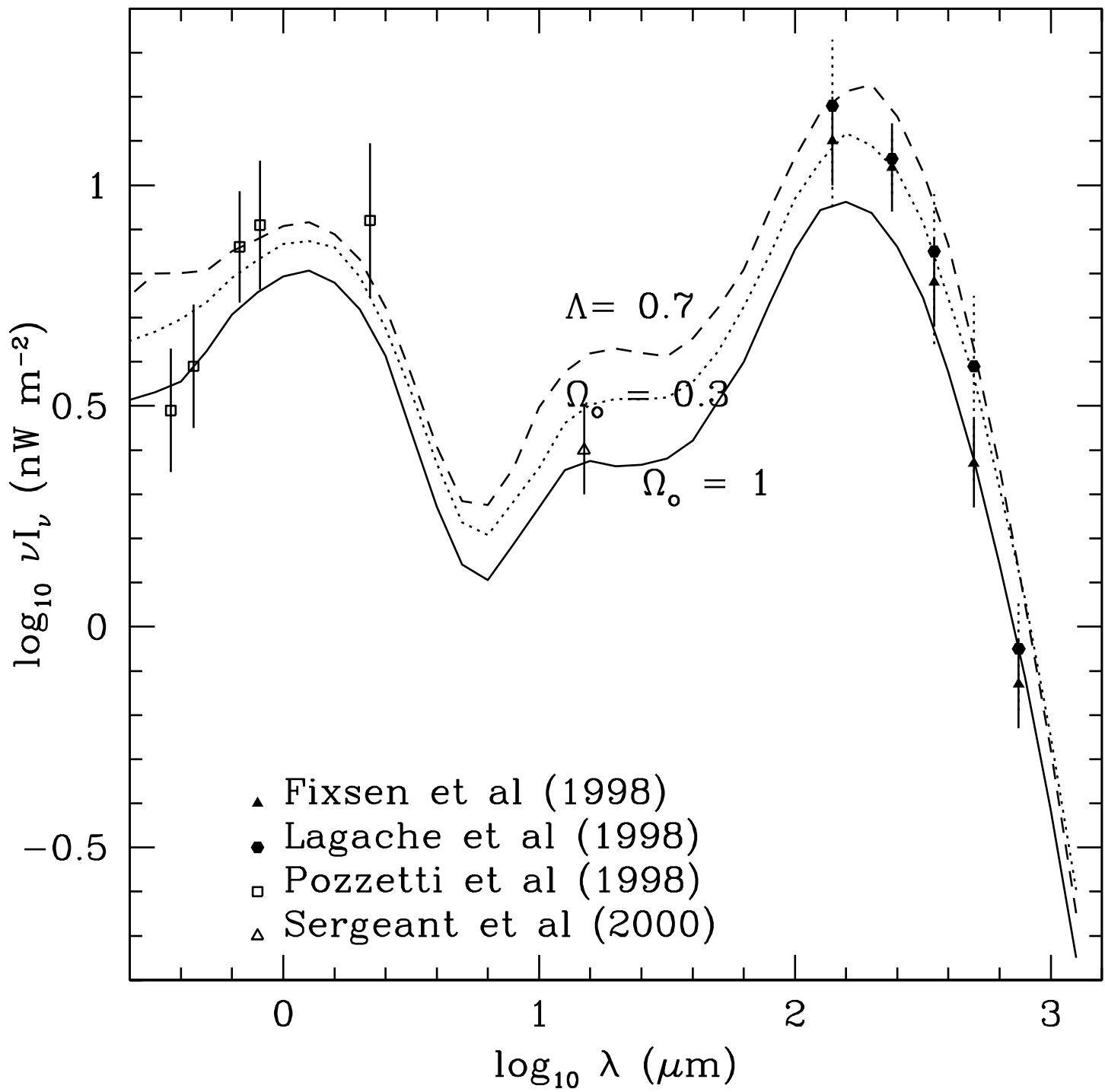
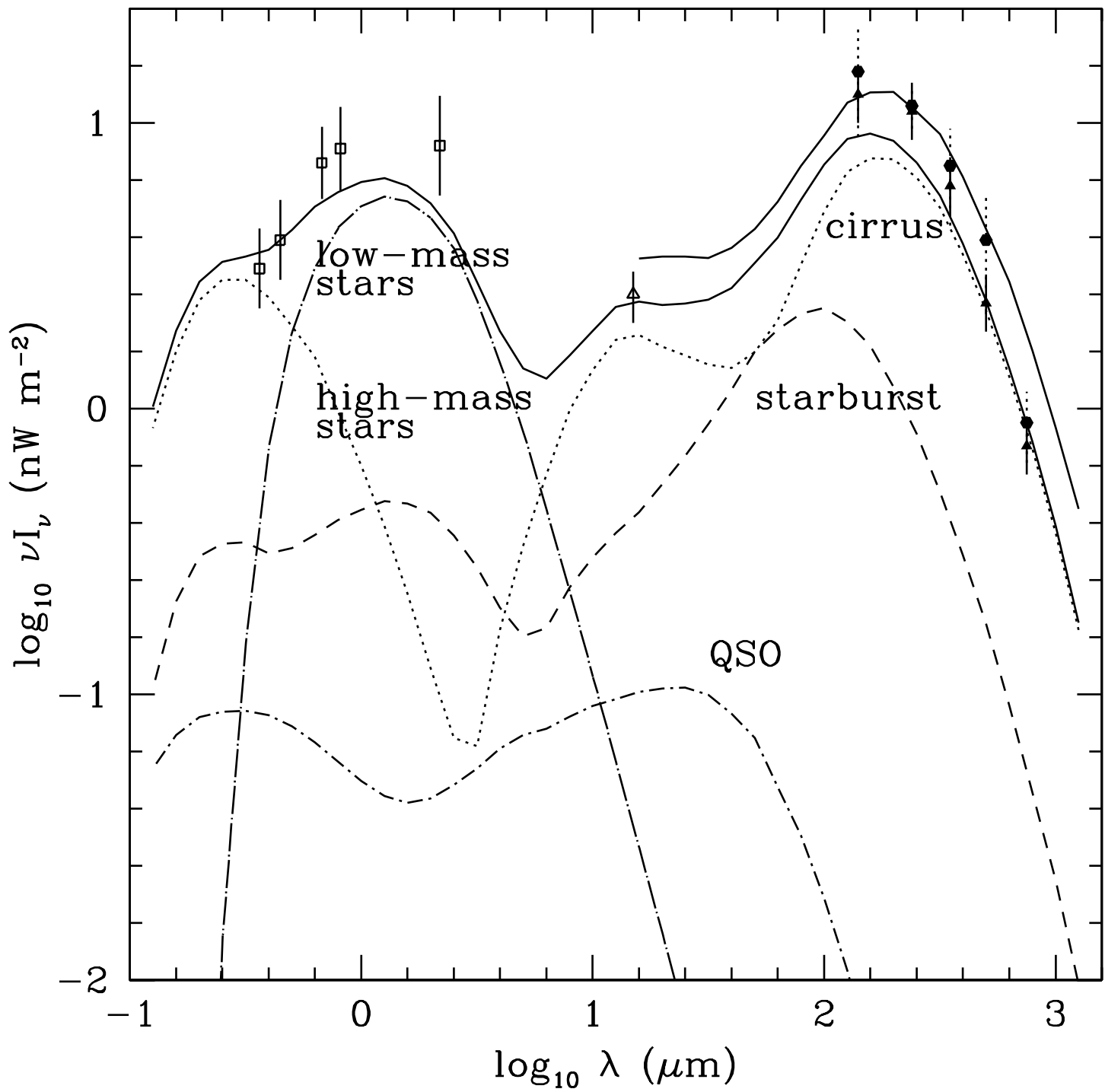
\end{document}